\documentclass[apj]{emulateapj}

\usepackage{multirow}
\usepackage{graphicx}
\usepackage{color}
\usepackage[pdftex,
        colorlinks=true,
        urlcolor=blue,       
        filecolor=blue,     
        citecolor=blue,       
        pdfpagemode=None,
        bookmarksopen=true]{hyperref}
        
\usepackage{natbib,twoopt}
\usepackage{amsmath}

\def\HI{\ion{H}{1} }
\def\CI{\ion{C}{1} }
\def\HII{\ion{H}{2} }
\def\CII{\ion{C}{2} }

\newcommand{\sio}{\ensuremath{\rm SiO}}

\newcommand{\Ms}{\ensuremath{M_{\sun}}}

\newcommand{\Mloss}{\ensuremath{M_{\sun}\,\rm pc^{-2}\,Gyr^{-1}}}
\newcommand{\cms}{\ensuremath{\rm cm^{-2}}}
\newcommand{\cmc}{\ensuremath{\rm cm^{-3}}}
\newcommand{\kms}{\ensuremath{\rm km\,s^{-1}}}

\newcommand{\ddr}[1]{{{\rm d}\, {#1} \over{\rm d}\,r}}

\newcommand{\dd}[1]{{\rm d} {#1} }

\usepackage{txfonts}
\usepackage{ulem}

\begin{document}
   \title{Can star cluster environment affect dust input from massive AGB stars?}

 \shorttitle{Massive AGB stars in star clusters}

   \author{Svitlana~Zhukovska\altaffilmark{1,2}, Mykola~Petrov\altaffilmark{3}, and  Thomas~Henning\altaffilmark{2}}
   \altaffiltext{1}{Max Planck Institute for Astrophysics, Karl-Schwarzschild-Str. 1, D-86748 Garching, Germany}
   \altaffiltext{2}{Max Planck Institute for Astronomy, K\"onigstuhl 17, D-69117 Heidelberg, Germany}
   \altaffiltext{3}{Institute for Astrophysics, University of Vienna, T\"urkenschanzstrasse 17, A-1180 Vienna}

               
  \begin{abstract}
We examine the fraction of massive asymptotic giant branch (AGB) stars remaining bound in their parent star clusters and the effect of irradiation of these stars by  intracluster ultraviolet (UV) field. We employ a set of N-body models of  dynamical evolution of star clusters rotating in a galactic potential at the solar galactocentric radius.  The cluster models are combined with stellar evolution formulae, a library of stellar  spectra, and simple models for SiO photodissociation in circumstellar environment (CSE). The initial stellar masses of clusters are varied from $50\Ms$ to $10^{5}\Ms$. Results derived for individual clusters are combined using a mass distribution function for young star clusters. We find that about 30\% of massive AGB stars initially born in clusters become members of the field population, while the rest evolves in star clusters. They are irradiated by strong intracluster UV radiation resulting in the decrease of the photodissociation radius of SiO molecules, in many stars down to the dust formation zone. In absence of dust shielding, the UV photons penetrate in the CSE deeper than $10R_*$ in 64\% and deeper than $2 R_*$  in 42\% of all massive AGB stars. If this suppresses following dust formation, the current injection rate of silicate dust from AGB stars in the local Galaxy decreases from $2.2 \times 10^{-4}\Mloss$ to $1.8 \times 10^{-4}\Mloss$ at most. A lower revised value of 40\% for the expected fraction of presolar silicate grains from massive AGB stars is still high to explain the non-detection of these grains in meteorites. 
\end{abstract}


   \maketitle


\section{Introduction}
Circumstellar shells of low- and intermediate-mass stars ($0.8\Ms<M<8\Ms$\footnote{We do not consider super AGB stars and assume that the upper limit is determined by the maximum mass of stars that develop an electron-degenerate C−O core and end their life as white dwarfs \citep[but see][]{Doherty:2013fb}.}) at the thermally-pulsing AGB are known sites of dust condensation. Some grains from AGB stars, which finished their interstellar journey in material forming the solar system 4.6~Gyr ago, have been preserved in meteorites. Presolar origin of these grains (initial masses and metallicities of their parent stars)  can be ascertained from anomalous isotopic ratios of the major and many trace elements characterising stellar nucleosynthesis  \citep{Hoppe:2000ds, Nittler:2003jv, Dorschner:2010ks, 2014mcp..book..181Z}.

Silicates and oxide grains are condensed in oxygen-rich stellar winds of AGB stars. The third dredge-up process mixes the carbon-rich ashes from He nuclear burning to the convective envelope turning a star into a carbon star. This process becomes more efficient with increasing stellar mass, therefore only low-mass AGB stars in the mass range of 1-1.5~\Ms\ are prominent sources of oxygen-rich dust.  However, massive AGB stars with initial mass from 3.5 -- 5~\Ms\ up to 8~\Ms, depending on model details and metallicity \citep{Busso:1999p5975, Marigo:2007dc}, convert the dredged-up carbon into $^{14}$N as a result of the hot-bottom burning process. There is recent spectroscopic evidence that hot-bottom burning is active from the first thermal pulses through their final superwind phase \citep[e.g.,][]{GarciaHernandez:2013ko, Justtanont:2012hy}. 

Theoretical models of dust condensation in stellar winds of evolved stars indicate that massive AGB stars should  produce significant amount of oxygen-rich dust \citep[e.g.][]{Gail:2010fs, Ferrarotti:2006p993, Ventura:2012cs, Nanni:2013wb}. Models of the lifecycle of grains of different origins show that the oxygen-rich grains from massive AGB stars are expected to be an abundant  component of the presolar grain population comparable to that from low-mass stars \citep{Gail:2009p512}. The fact that grains are efficiently condensed in shells of AGB stars experiencing hot-bottom burning is supported by spectroscopic observations of oxygen-rich dust-enshrouded stars \citep[e.g.,][]{VanLoon:2005ik}. The chemical composition of these stars is altered by the hot-bottom burning process leading to subsolar values of $\rm ^{12}C/^{13}C$ and $\rm ^{18}O/^{17}O$ isotope ratios \citep{Boothroyd:1995es}, which should be possible to detect in presolar grains. 
However, no oxygen-rich grain with signatures of hot-bottom burning has been conclusively identified  in meteorites so far. The entire population of presolar oxygen-rich grains appears to originate from low-mass AGB stars \citep[][]{Vollmer:2008kc, Nittler:2008bp, Iliadis:2008id, Nittler:2009ck, Hoppe:2010tg, Palmerini:2011eh}. To shed light at the origin of this discrepancy between theoretical models and findings from presolar grain studies, we consider additional processes that can potentially reduce the global dust input from intermediate-mass stars.

The total contribution of massive AGB stars to the Galactic dust budget depends on the adopted dust yields, i.e. on the amount of  dust condensed in the stellar wind over the whole AGB evolution of a star. Dust yields as a function of the initial stellar mass and metallicity have been derived over the past decade from extensive calculations combining stellar evolution at AGB stars, stellar wind and dust condensation models  \citep[e.g.,][]{Ferrarotti:2006p993, Ventura:2012cs, Nanni:2013wb}. Such models consider single isolated AGB stars under standard interstellar conditions. Nucleation and growth of dust particles in outflows of these stars is assumed to be independent of their environment. This is justified for the irradiation of CSE by the standard interstellar radiation field, because the dust formation zone is located much deeper in the shell than the photodissociation region \citep[e.g.,][]{Jura:1981dk, Glassgold:1987wj, Mamon:1987gv, Glassgold:1996kg}. 
This may not be the case for evolved stars in young star clusters.
Unlike low-mass stars, intermediate-mass stars are more likely to be members of their parent star clusters during their AGB evolution. 

The majority of stars in the mass range $4\Ms \leqslant M < 8\Ms$ is formed in massive star clusters, which survive cloud dispersal and become open clusters, as will be shown below. The dissolution time of open star clusters exceeds $\sim 200$~Myr, the lifetime of a 4~\Ms{} star \citep{2009A&A...495..807K, 2009MNRAS.392..969J, Baumgardt:2003dl}. The circumstellar environment (CSE) of massive AGB stars in young star clusters are subject to irradiation by the intracluster UV field from main sequence cluster stars  of spectral types B and early A. Strong UV irradiation of CSE may affect the dust condensation process. \cite{Beck:1992p6642} investigated dust condensation in stellar winds of stars with ionising radiation of chromospheric origin and found that UV photons are able to reduce and, for certain field strengths, completely suppress the nucleation process.  
If the amount of dust condensed in outflows of massive AGB stars in star clusters is affected by external UV radiation, the net dust input from these stars will be reduced. 

In order to assess the importance of cluster environment on CSE of massive AGB stars, first of all we need to estimate the fraction of intermediate stars in the local Galaxy that evolve in their parent star clusters. To this end, we employ numerical simulations of the dynamical evolution of star clusters in a mass range of 50--$10^5\Ms$ coupled with stellar evolution prescriptions \citep{2009A&A...495..807K, 2009MNRAS.392..969J, 2010A&A...524A..62E}. 
Next, for the massive AGB stars that remain bound in the cluster, we calculate the ultraviolet radiation field created by the neighbouring main sequence cluster members as a function of cluster mass. For a simple estimate of the impact of the intracluster radiation field on the circumstellar shells of AGB stars in cluster, we consider photodissociation of the SiO molecule. The SiO molecules are the basic building block for silicates, the most widespread dust species in oxygen-rich environment \citep{Henning:2010p7233, Gail:2014tu}. They are formed in the stellar photosphere, and in the shell their abundance decreases due to two processes: depletion on grains in the dust condensation zone and then photodissociation caused by interstellar UV photons in the outer regions. We calculate the photodissociation radius of SiO molecule for each massive AGB stars in the model clusters and compare these radii to the position of the dust formation zone. 

The structure of our paper is as follows. Section~\ref{sect:Mod} describes a suite of dynamical models of star cluster evolution, size-frequency distribution of star clusters and our main assumptions in calculations of the UV fluxes irradiating circumstellar shells of massive AGB stars. Results of model calculations of dynamical evolution of model star clusters are presented in Sect.~\ref{sec:results}. We show that about 70 \% of massive AGB stars initially born in clusters remain cluster members. For these stars, we investigate the location of the photodissociation radii of SiO molecule with respect to the dust formation zone in their CSE. Dependence of our main results on the initial mass--radius relation for star clusters is presented in Sect.~\ref{sec:dependIC}. Discussion and concluding remarks are given in Sect.~\ref{sec:discussion}.

\section{Model}\label{sect:Mod}

\subsection{Simulations of star cluster evolution} \label{sec:SC-evolution}
The dynamical evolution of star clusters (SCs) is modelled as pure N-body systems of gravitationally bound stars
rotating in a galactic potential at the solar galactocentric radius of 8.5\ kpc. The simulations were carried out with the direct $N$-body code $\phi$GRAPE \citep{2007NewA...12..357H} on high-performance computing
clusters with accelerator hardware GRAPE or GPU selected by appropriate libraries.  These models were introduced
and applied to individual SCs in a series of papers \citep{2009A&A...495..807K, 2009MNRAS.392..969J, 2010A&A...524A..62E}. In the following, we briefly summarise the main model parameters used in our calculations
and refer to the original publications for a detailed description.

\begin{table*}[!t]
\caption{Initial parameters of the N-body models}
\label{tab:InitialParamNbody}
\vspace{3mm}
\begin{tabular}{r c c c c c c}
\hline\hline
$N_*$ & $N_\mathrm{sim}$ & $M_\mathrm{cl}$ & $R_\mathrm{hm}$ (MRR) & $R_\mathrm{hm}$ (TRR) & $N_{4 \leqslant m /M_{ \odot} < 8}$ & $N_{2 \leqslant m/M_{ \odot}  < 8}$ \\
      &                  &  ($\Ms$)          & (pc)          & (pc)          &    & \\
\hline
109     & 4 & $50$              &  0.25 & 0.75 & 1,2,1,1         & 3,3,4,5     \\
153     & 4 & $70$              &  0.35 & 0.84 & 1,2,2,2         & 5,5,6,6     \\
218     & 4 & $10^{2}$          &  0.44 & 0.95 & 1,2,2,2         & 7,4,7,8     \\
437     & 4 & $2 \times 10^{2}$ &  0.54 & 1.20 & 3,2,5,5         & 15,9,17,15  \\
1094    & 4 & $5 \times 10^{2}$ &  0.85 & 1.63 & 11,12,10,9      & 30,28,40,27 \\
1532    & 4 & $7 \times 10^{2}$ &  1.04 & 1.82 & 16,21,13,12     & 50,51,47,54 \\
2189    & 4 & $10^{3}$          &  1.29 & 2.05 & 19,25,21,23     & 60,69,66,77 \\
4379    & 4 & $2 \times 10^{3}$ &  1.81 & 2.58 & 36,49,48,46     & 130,155,140,142 \\
10947   & 4 & $5 \times 10^{3}$ &  2.85 & 3.50 & 100,118,105,111 & 364,373,345,351     \\
15326   & 4 & $7 \times 10^{3}$ &  3.30 & 3.92 & 164,132,142,149 & 519,510,492,492     \\
21895   & 1 & $10^{4}$          &  3.99 & 4.41 & 203             & 716         \\
43791   & 1 & $2 \times 10^{4}$ &  5.69 & 5.56 & 415             & 1454        \\
109476  & 1 & $5 \times 10^{4}$ &  8.96 & 7.55 & 1025            & 3620        \\
153268  & 1 & $7 \times 10^{4}$ & 10.62 & 8.44 & 1482            & 5096        \\
218955  & 1 & $10^{5}$          & 12.66 & 9.51 & 2059            & 7272        \\
\hline
\end{tabular}
\\[3mm]
\textbf{Notes.} The first column gives the total number of stars, followed by the number of simulation runs, the initial cluster mass, the initial half-mass radius for the mass-radius relation and that for the tidal radius-mass relation. The last two columns give the number of massive AGB stars and the number of irradiating stars in each simulation.
\end{table*}

\subsubsection{Initial mass--radius relation}
We consider the evolution of low- and intermediate-mass stars in clusters, after the last massive stars ended their life as supernova, corresponding to the cluster ages of 40~Myr. Initially star clusters are embedded in their parent molecular clouds which are completely cleared off by stellar feedback by the onset of our simulations. Modelling of embedded cluster evolution and their emergence from parent molecular clouds requires including of gas dynamics or special algorithms accounting for the influence of gas on star dynamics as discussed in \cite{Pelupessy:2012ca} and \cite{Baumgardt:2007kg}. These authors showed that the dynamical friction drives early mass segregation resulting in preferential retention of  stars with mass above $2\Ms$ in clusters. Therefore we neglect possible ejection of stars during the early cluster evolution before the onset of our simulations.

For the initial distribution of stars in the clusters, we adopt commonly used not-rotated King models \citep{King:1966id} with a central potential of $W_0 = 6.0$ \citep[][]{1999MNRAS.302...81E}. In this case, the initial mass distribution is given by three parameters: the initial cluster mass, concentration parameter and scale radius. Observed shapes of SCs as well as their sizes for a given cluster mass vary in a large range  and can poorly constrain the initial conditions for simulations \citep{Larsen:2004gf, PortegiesZwart:2010kc}. This is due to the fact that they are influenced by a number of physical processes such as cluster formation, internal processes (self-gravitation, binary fraction, rotation), external forces (tidal field of the galaxy), encounters with giant molecular clouds \citep[][]{Scheepmaker:2007bt, Piskunov:2008hp, Kharchenko:2009fs}. Importance of these processes for the initial conditions of SC models is reviewed by \cite{Kroupa:2008hs}. To minimize the number of free parameters, we adopt a relation between the initial mass and radius 
\begin{equation}
\label{eq:MRR}
	R_\mathrm{cl} \approx 50 \sqrt{ M_\mathrm{cl} / 10^6 \Ms } \, (\mathrm{ pc})
\end{equation}
derived from observations of molecular clouds and clumps in the Milky Way \citep{Rivolo:1988ha, Larson:1981vv}. 
This mass-radius relation (MRR) has been used 
to set up the initial conditions for simulations of open clusters in the Milky Way \citep{Adams:2006cl, 2008ApJ...675.1361F, 2009A&A...495..807K}. It has been also applied in theoretical studies of tidal tail clumps  of SCs \citep{2009MNRAS.392..969J} and for calibration of radii and masses of observed clusters with simulations \citep{2010A&A...524A..62E}.
The initial half-mass radius of each cluster was adjusted so that the half-mass radius of a King model was initially equal to 80 per cent of a scaling radius and inside of which 60 per cent of cluster mass is enclosed.

Given that there is no strong observational constraints or a universal relation for the initial conditions of simulated clusters, we test the dependence of our results on the adopted initial cluster mass--radius relation and perform two additional sets of simulations. We follow an alternative method to set up the initial conditions assuming that the initial cluster radius is determined by the tidal field of the Galaxy and is given by \citep{1983AJ.....88..338I, Baumgardt:2003dl, Lamers:2013bo}
	\begin{equation}
	\label{eq:TRR}
		R_\mathrm{tidal} = (0.5 GM_\mathrm{cl} / V_\mathrm{G}^2 )^{1/3} R_\mathrm{G}^{2/3},
	\end{equation}
where $M_\mathrm{cl}$ is the cluster mass,  $R_\mathrm{G}$ the galactocentric distance of the cluster and $V_\mathrm{G}$ the circular velocity of the Galaxy. This set of models is denoted as TRR.  The half-mass radii for TRR and MRR models are shown in Table~\ref{tab:InitialParamNbody}.

Additionally, we perform a simulation run with the constant initial radius for all clusters (CRR models) to include the case of a weak mass-radius relation suggested by some observations of galactic and extragalactic star clusters \citep[][]{Larsen:2004gf, Scheepmaker:2007bt}. We adopt a value of the initial half-mass radius of 2~pc \citep{Lamers:2013bo}, consistent with the range of values from observations. In the following, we use the MRR models as the reference models and refer to them in the analysis of the results, unless stated otherwise. Comparison of the main results for the TRR and CRR models with the reference models are given in Sect.~\ref{sec:dependIC}.

\subsubsection{Stellar mass distribution in SCs}
The stellar mass in clusters is distributed following the initial mass function (IMF) from \citet{2002Sci...295...82K} 
where only stars in the mass range of $0.08 \leqslant m \leqslant 8.0~\Ms$ are studied. Stellar evolution of stars in clusters 
(stellar luminosities, radii, effective temperatures at various evolutionary phases) is described  accordingly to the analytical 
formulas from \citet{2000MNRAS.315..543H}. Binaries as well as multiple star systems were not included in our study. 
Possible encounters of SCs with giant molecular clouds as well as passages of spiral arm or disk shocking were neglected.

The initial mutual distribution of stars of different mass in clusters is still debated. 
There are many observational studies in support of the initial mass segregation (IMS) of stellar mass, i.e. concentration 
of more massive stars in the inner cluster region \citep[see][and references therein]{Lada:2003il, PortegiesZwart:2010kc}. 
However, observational evidence of mass segregation was questioned in \cite{Ascenso:2009ky}, who pointed at difficulties to 
differentiate between segregation and sample  incompleteness effects. There are observations that indicate that very low-mass 
stars associated with a cluster are distributed homogeneously in a volume that is much larger than the core of a cluster 
\citep{Kumar:2007iv}. Given these uncertainties in the initial mass distribution in star clusters, we run two sets of models: 
with and without initial mass segregation to study how it affects the evolution of massive AGB stars.
Initial conditions for clusters with segregation are derived accordingly to the procedure described by \citet{2008ApJ...685..247B}. 
It invokes a random number generator to draw distributions of initial stellar masses, positions, and velocities of stars in clusters.

\subsubsection{Initial parameters of model clusters}
We consider the dynamical evolution of  15 clusters with and without initial mass segregation  resulting in 30 setups of simulation models. 
Initial masses, radii and numbers of stars for each cluster are listed in Table~\ref{tab:InitialParamNbody}. All clusters are moving on circular orbits in the Galaxy represented by an analytic background potential described 
by Plummer-Kuzmin models \citep{1975PASJ...27..533M} with circular velocity $V_\mathrm{G} = 233 \, \mathrm{km \, s^{-1}}$. 
Model SCs are assumed to have solar metallicity.
For small clusters, the number of stars within a given mass range depends on randomized discretization of the initial stellar mass 
over individual stars. Therefore, we generate several sets of initial conditions for star clusters with the initial 
masses $M_\mathrm{cl} \leqslant 7\times 10^3\Ms$ resulting in 90 simulation models in total for one set. The number of simulation runs for each cluster model is given in the second column 
of Table~\ref{tab:InitialParamNbody}. The table also shows the total numbers of stars with mass $m \geqslant 4\Ms$, 
which experience hot bottom burning during TP-AGB evolution, and with $m \geqslant 2 \Ms$, which dominate the intracluster 
UV radiation field, in each simulation run. 
Accordingly to \citet{2013MNRAS.434...84W} \citep[see also][]{2013pss5.book..115K} the lowest mass of a star cluster to 
host at least one massive TP-AGB star is $\sim 50\Ms$. Therefore this value is adopted here for $M^{\min}_\mathrm{cl}$. We adopt an upper mass limit of model clusters  $M^{\max}_\mathrm{cl}=10^5\Ms$, appropriate for the Milky Way \citep[][and references therein]{Kruijssen:2014if}. Star clusters of higher mass in the Galaxy belong to metal-poor $\gtrsim 10$~Gyr old globular cluster systems. Given short lifetimes of grains in the ISM \citep{Slavin:2015in}, dust from massive AGB stars from globular clusters could not survive until the solar system formation and is excluded from our consideration.

Relevant information about all stars from the simulation runs (e.g., their coordinates, masses, luminosities, temperatures) is stored in snapshots with a time step of  $\sim 0.1$~Myr. It is thus ensured that each massive TP-AGB star appears in several snapshots. 

\subsection{Fraction of massive AGB stars in clusters}\label{sec:ModFracAGBstars}
 In order to combine the results derived for individual star clusters of different masses to estimate the total fraction of massive AGB stars evolving in clusters, we need to know the cluster mass distribution function for the onset of simulations corresponding to the cluster age of 40~Myr. We assume that most of stars are formed in clusters embedded within giant molecular clouds (\citealt{Lada:2003il}; but see also \citealt{Kruijssen:2012bs, Adamo:2015bn}). The distribution function of embedded star clusters can be measured observationally by infrared star counts. However, only the most massive ($M_\mathrm{cl} > 500 \Ms$) embedded clusters survive and become stable, open clusters \citep{Lada:2003il, Bonatto:2011ev}. Since by the age of 40~Myr all clusters should emerge from their birth places, we can adopt the mass distribution function of embedded star clusters, with a correction for the infant mortality of low-mass star clusters as described below.

\subsubsection{Embedded star cluster mass function}
 Observations indicate that the mass-frequency distribution of young embedded star clusters $\dd n_\mathrm{emb}(M_\mathrm{cl})/\dd M_\mathrm{cl}$ follows 
a universal power law \citep[][]{Kennicutt:1989go, McKee:1997vf, Lada:2003il, Larsen:2010il, PortegiesZwart:2010kc}:
\begin{equation}
     \dd{n_\mathrm{emb}(M_\mathrm{cl})}/\dd{M_\mathrm{cl}} \propto M_\mathrm{cl}^{-2}, \textrm{\hspace{.5cm}}  M_\mathrm{cl}^{\min} \leqslant M_\mathrm{cl} \leqslant M_\mathrm{cl}^{\max} ,
     \label{eq:DFSC}
\end{equation}
where  $\dd{n_\mathrm{emb}(N_\mathrm{cl})}$  is the number of embedded star clusters with the masses
between $M_\mathrm{cl}$ and $M_\mathrm{cl} + \dd M_\mathrm{cl}$, $M_\mathrm{cl}^{\min}$ and $M_\mathrm{cl}^{\max}$ are the lowest and highest star cluster masses. As discussed above, we assume $M_\mathrm{cl}^{\min}= 50\Ms$ and $M_\mathrm{cl}^{\max}=10^5\Ms$, respectively.

The total number of AGB stars formed in all clusters with the masses below $M_\mathrm{cl}$ is
\begin{equation}
   \label{eq:IntegNum}
 	N^\mathrm{tot}_\mathrm{AGB} (M_\mathrm{cl}) = \int_{M_\mathrm{cl}^{\min}}^{M_\mathrm{cl}} N_\mathrm{AGB}(M) \frac{\dd{n_\mathrm{emb}(M)}}{\dd{M}}  \dd{M}.
\end{equation}
where $N_{\rm AGB}(M)$ is the initial number of AGB stars in a cluster with  the mass $M$. 
 
The cumulative number of AGB stars formed in a cluster with mass $M<M_\mathrm{cl}$ normalised to the total number of these stars is easily derived from Eq.~(\ref{eq:IntegNum}) 
\begin{equation}
	\label{eq:CumNum}
	\begin{split}
	f_\mathrm{AGB}^{\rm cum} (M_\mathrm{cl})  =  \int_{M_\mathrm{cl}^{\min}}^{M_\mathrm{cl}} N_\mathrm{AGB}(M)  \frac{\dd{n_\mathrm{emb}(M)}}{\dd{M}}\ \dd{M}  \, \Big{/} \,  \\
\int_{M_\mathrm{cl}^{\min}}^{M_\mathrm{cl}^{\max}} N_\mathrm{AGB}(M)  \frac{\dd{n_\mathrm{emb}(M)}}{\dd{M}} \dd{M}.
\end{split}
\end{equation}

\begin{figure}[!tb]
	\includegraphics[width=0.5\textwidth] {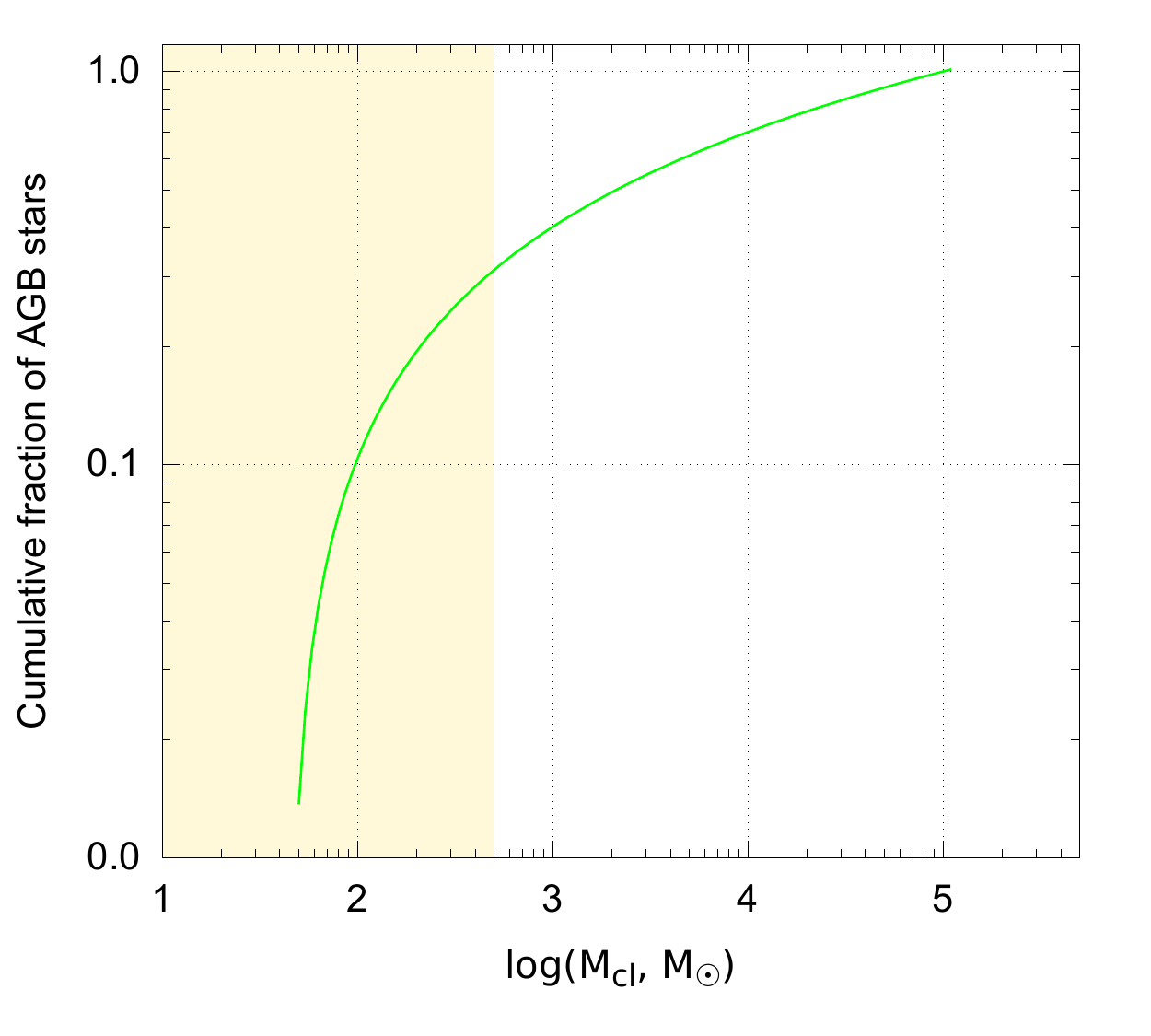}
	\caption{Cumulative distribution of progenitors of massive TP-AGB stars (in all clusters with mass below $M_{\rm cl}$ ) normalised to the total number of these stars in all clusters calculated using the embedded cluster mass function. The shaded area marks the initial mass of clusters, which emerge mostly unbound from their parental molecular clouds, accordingly to \cite{Lada:2003il}.}
	\label{fig:tpagb-sc}
\end{figure}

Figure~\ref{fig:tpagb-sc} shows the cumulative distribution of stars with initial masses $4 \Ms \leqslant M < 8 \Ms$, which are progenitors of massive TP-AGB stars, as a function of initial cluster mass for the distribution function of young clusters given by Eq.~(\ref{eq:DFSC}). It is derived for the adopted Kroupa IMF and cluster mass range. 

Because of gradual dissolution of SCs, the actual values of bound AGB stars $N^b_{\rm AGB}(M_\mathrm{cl})$ derived from simulations are lower. Additionally, the mass distribution function of embedded star clusters has to be corrected for the dissolution of low-mass clusters upon dispersal of molecular clouds as described below.

\subsubsection{Dissolution of low-mass clusters}\label{sec:Dissol}
Numerical models in this work consider evolution of clusters shortly after emergence from molecular clouds, therefore we include two sources of the field population of massive AGB stars separately: (i) members of low-mass clusters ($M_\mathrm{cl} < 500 \Ms$) which did not survive dispersal of molecular clouds, and (ii) stars lost by bound clusters in the process of their dynamical evolution. The latter is derived from our dynamical N-body models as a function of initial cluster mass. 

The fraction of star clusters that survive as bound systems up to Pleiades age was assessed by \cite{Lada:2003il} by comparison of the numbers of embedded and open star clusters from observations. They found that majority (90--95\%) of embedded star clusters with initial mass $M_\mathrm{cl}  \lesssim 500 \Ms$ emerge from molecular clouds as unbound systems and contribute their members to the field population. The contribution of clusters in this mass range to the total number of progenitors of massive TP-AGB stars is about $30 \%$, while about a half of progenitors of massive AGB stars are formed in clusters with the mass above $2 \times 10^3 \Ms$ (Fig.~\ref{fig:tpagb-sc}). 

Dynamical models which consider embedded evolution of star clusters in molecular clouds find that the survival of embedded star clusters depends on many physical parameters such as gas expulsion timescale, star formation efficiency, and impact of an external tidal field \citep{2012MNRAS.427.1940P, 2011MNRAS.413.1899P}. Nevertheless, the slopes of the mass functions of embedded or young star cluster and gas-free clusters are identical \citep{Dowell:2008ba, Oey:2004bs, Lada:2003il}. For a simple estimate of the contribution of clusters $M_\mathrm{cl} \leqslant 500 \Ms$  to the field population, we assume that initially these SCs were formed accordingly to the mass distribution function $ \dd{n_\mathrm{emb}(M)}/ \dd{M}$ given by Eq~(\ref{eq:DFSC}) as their higher mass counterparts, but 90\% of these clusters do not survive cloud dispersal based on the lower limit derived by \cite{Lada:2003il} and more recently by \cite{Bonatto:2011ev}. Therefore, for the onset of our simulation, we adopt
the mass distribution function of young star clusters $\dd{n_\mathrm{corr}(M)}/ \dd{M}$ corrected for the dissolution of low-mass clusters by multiplication of $ \dd{n_\mathrm{emb}(M)}/\dd{M}$ by 0.1 for $M_\mathrm{cl} \leqslant 500 \Ms$.

The total number of bound AGB stars is
\begin{equation}
	\label{eq:NumBound}
	N^\mathrm{bound}_\mathrm{AGB} = \int_{M_\mathrm{cl}^{\min}}^{M_\mathrm{cl}^{\max}} N^\mathrm{b}_\mathrm{AGB}(M)  \frac{\dd{n_\mathrm{corr}(M)}}{\dd{M}} \dd{M} \,
\end{equation}
where $N^\mathrm{b}_\mathrm{AGB}(M)$ is the number of bound stars in a SC with mass $M$ derived from numerical simulations of dynamical SC evolution. The total number of bound AGB stars residing in star clusters with $M<M_\mathrm{cl}$ relatively to the number of all AGB stars therefore equates to
\begin{equation}
	\label{eq:CumNumCorrected}
	f_\mathrm{AGB}^\mathrm{cum} (M_\mathrm{cl})  =  \frac{1}{N^\mathrm{tot}_\mathrm{AGB}} \int_{M_\mathrm{cl}^{\min}}^{M_\mathrm{cl}}
	 N^\mathrm{b}_\mathrm{AGB}(M)  \frac{\dd{n_\mathrm{corr}(M)}}{\dd{M}}\ \dd{M} ,
\end{equation}
where $N^\mathrm{tot}_\mathrm{AGB}$ is given by Eq.~(\ref{eq:IntegNum}) for $M_\mathrm{cl}=M_\mathrm{cl}^{\max}$.
 

\subsection{Intracluster UV radiation field}
\begin{figure}[tb]
	\includegraphics[width=0.5\textwidth]{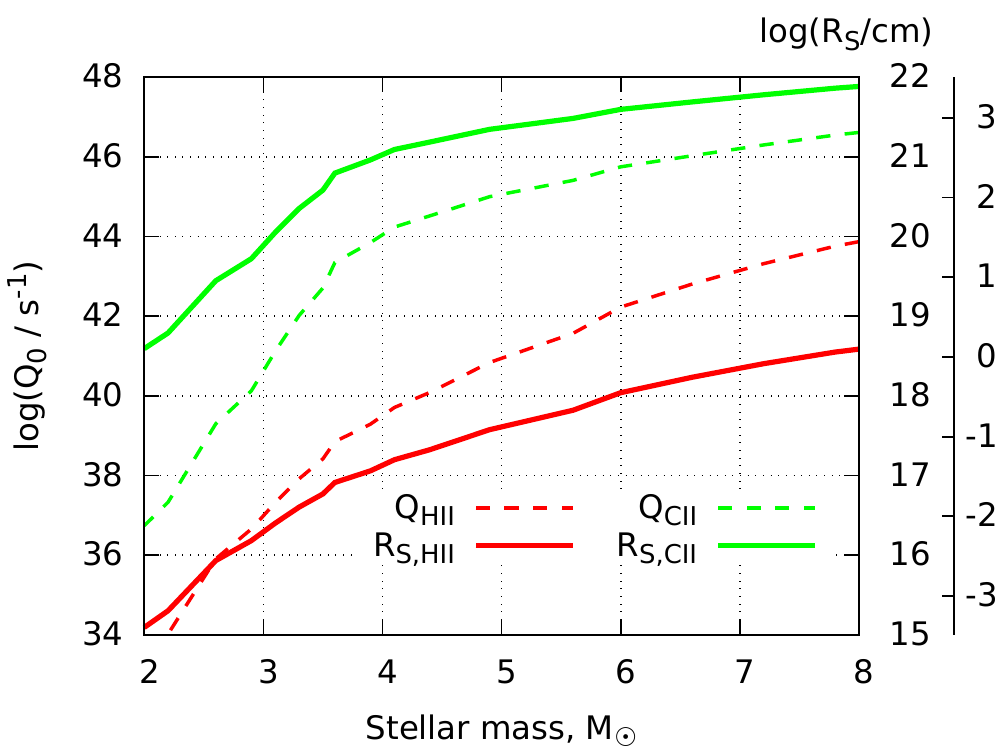}
	\caption{Number of ionising photons for H and C atoms emitted per unit time as a function of initial stellar mass (left axis) and corresponding radii of ionised zones (right axis) calculated for the value of ambient gas density of $n_{\rm H} = 1\ \cmc$. The second axis on the right shows $\log R_s$ in pc.}
	\label{fig:ion-flux}
\end{figure}

Circumstellar envelopes of evolved stars in young star clusters are irradiated by UV radiation emitted by the main sequence cluster members of spectral type A and late B. In older clusters, thermal radiation from white dwarfs becomes an important source of UV photons, which is probably responsible for the ionisation and removal of intracluster medium \citep{McDonald:2015jr}.
The strength of the intracluster UV radiation field due to main sequence stars is expected to quickly decrease over a few hundred Myr as the cluster turn off point shifts towards stars of lower mass, since amount of emitted UV radiation is a strong function of stellar mass. It is illustrated in Fig.~\ref{fig:ion-flux} showing the total number of ionising photons for \HI and \CI atoms emitted per unit time $Q_0$ as a function of initial mass for intermediate-mass stars.  $Q_0$ is derived by integration of the stellar surface flux $F_\nu$ 
\begin{equation}
\label{eq:IonRate}
Q_0 = \int_{\nu_{\min}}^{\nu_{\max}} {\frac{4\pi R_*^2 F_\nu}{h \nu}  \dd \nu},
\end{equation}
where $R_*$ is the stellar radius, and $\nu_{\max}$ and $\nu_{\min}$ correspond to the wavelength of 91.178~nm and 240~nm (5.17~eV) for \HI and to 91.178~nm and 110.11~nm (11.26 eV) for \CI, respectively.

Intermediate-mass main sequence stars in clusters are capable to ionise the intracluster medium. Their radiation is hardly attenuated within the cluster owing to the observed lack of gas and dust in the intracluster medium \citep{Bastian:2014ea}. Figure~\ref{fig:ion-flux}  shows the radii of Str\"omgren spheres, i.e. zones of ionised \HII and \CII around the central star, for the homogeneous intracluster gas density of 1~\cmc, similar to the density of the ambient ISM. Although the sizes of ionised regions around intermediate-mass stars are small compared to those around massive stars, they are comparable to the sizes of low-mass clusters (Table~\ref{tab:InitialParamNbody}). 
After the onset of mass loss from evolved stars in clusters, the density distribution in the vicinity of these stars will follow $r^{-2}$ radial dependence and only outer layers will be ionised by the extreme UV radiation. 

The strength of the UV field is customary characterised relatively to the standard ISRF \citep{Rollig:2007p6883}
\begin{equation}
\label{eq:chi}
 \chi = \int_{91.2\rm nm}^{240\rm nm} \lambda u_{\lambda,i} \dd{\lambda}  \, \Big{/} \,
  \int_{91.2\rm nm}^{240\rm nm}  \lambda u^{\rm ISRF}_\lambda \dd{\lambda},
\end{equation}
where $u_\lambda = 4\pi/c J_\lambda$  the spectral photon energy density, $u_\lambda^{\rm ISRF}$ is the the spectral photon energy density of the interstellar radiation field \citep{Draine:1978p5783}. The mean radiation field intensity in the location of an AGB star is approximated as
\begin{equation}
\label{eq:MeanIntens}
  {4\pi} J_\lambda =   \sum_{i=1}^{n_{\rm MS}}  \Omega_{*}^i I_{\lambda, *}^i + 4 \pi J_{ISRF} =  \sum_{i=1}^{n_{\rm MS}}  \frac{R_{*i}^2} {r_i^2 }  I_{\lambda, *}^i + 4 \pi J_{ISRF}
\end{equation}
where $n_{\rm MS}$ is the current number of main sequence stars with $m_* > 2\Ms$ in a cluster, $I_{\lambda, *}^i$ is the incident intensity from $i$th star, $R_i$ is its photospheric radius,  and $r_i$ is the distance from the central AGB star to $i$th star, and $\Omega_{*}^i $ is the solid angle subtended from the $i$th star. The last term on the right side of equation is the contribution from the interstellar radiation field. $I_{\lambda, *}^i$ is derived from the Eddington fluxes $H_\nu$ provided by a stellar spectral library
	\begin{equation}
 	I_\lambda = \frac{F_\lambda}{ \pi} = \frac{4c}{\nu^2} H_\nu 
	\end{equation}
The stellar fluxes are assigned to cluster stars accordingly to their temperature and surface gravity stored in the snapshots. We employ the BaSeL 3.01 semi-empirical library of stellar spectra for wavelength from 9 nm to 160~$\mu m$  \citep{Lejeune:1997gs, Lejeune:1998ed, Westera:2002hy}.

\subsection{Photodissociation radius of SiO molecules}\label{sec:ModIrrad}
SiO molecule is considered a prerequisite for formation of quartz and various silicate types of grains \citep{Henning:2010p7233, Gail:2014tu}.  
It is well established by photochemical models and supported by observations that the abundance of SiO molecules decreases in the CSE of AGB stars with radius by two processes: depletion on grains in dust condensation zone and dissociation by interstellar UV photons in outer regions \citep{Jura:1981dk, Huggins:1982gj, Huggins:1982cw, GonzlezDelgado:2003kn}. Such photochemical models usually assume a homogeneous dust distribution with dust attenuation properties similar to those  in the interstellar medium. In this case, dust shielding plays a crucial role in the chemistry of CSE, being able to protect molecules from the interstellar UV photons up to distances of $10^{16}-10^{17}$~cm \citep{Glassgold:1996kg}. 
 In this work, we evaluate how deep the UV photons can penetrate in the expanding shell of an AGB star, which is irradiated by the enhanced UV radiation prior to the onset of efficient dust condensation. We consider the sizes of the SiO envelopes around AGB stars in the model clusters and compare them to those irradiated by the ambient ISRF, and to the position of the dust formation zone. 

For a rough estimate, we adopt a toy model of the photodissociation of SiO molecule in a stationary spherically symmetric outflow, where the gas has been accelerated to terminal velocity $\upsilon_e$. In absence of other sinks and sources, the SiO fractional abundance $f_{\rm SiO}$ relatively to $\rm H_2$ changes due to photodissociation by UV photons as \citep{Jura:1981dk, Huggins:1982cw}
\begin{equation}
\label{eq:dfdt}
	\ddr{f_{\rm SiO}}=-\frac{k^{\rm pd}_{\rm SiO}}{\upsilon_e} \exp\left(-\frac{d_{\rm SiO}}{r} \right)
\end{equation}
where $d_{\rm SiO}$ is the shielding distance corresponding to the optical depth of 1 and $k^{\rm pd}_{\sio}$ is the unattenuated photodissociation rate of SiO molecules. For a monosize grain distribution, the shielding distance is 
\begin{equation}
\label{eq:dsio}
d_{\rm SiO} = 1.4 \frac{3 (Q/a)_{\sio}}{4 \rho_{\rm SiO}} \times \frac{\dot{M}_{\rm d}} {4\pi \upsilon_e},
\end{equation}
where $Q$ is the dust absorption efficiency, $a$ is the grain size, $ \rho_{\rm SiO}$ is the density of the solid material, $\dot{M}_{\rm d}$ is the dust mass-loss rate. The grain drift relatively to the gas was neglected. The value of dust mass loss is $\dot{M}_{\rm d} = D \dot{M}$, where $D$ is the dust-to-gas ratio in the stellar wind and $\dot{M}$ is the gas mass-loss rate from Eq.~(\ref{eq:dmdt}). 

The size of the SiO envelopes derived from observations agrees well with the photodissociation radius defined as the distance from the central star where the SiO abundance is decreased by $e$ times  \citep{GonzlezDelgado:2003kn}. The equation for the photodissociation radius is derived from Eq.~(\ref{eq:dfdt}) \citep{Jura:1981dk}
\begin{equation}
\label{eq:rpd}
	r_p=\frac{\upsilon_e}{k^{\rm pd}_{\rm SiO} E_2(d_{\rm SiO}/r_p)}.
\end{equation}
where $E_2(x)$ denotes the exponential integral. Equation~(\ref{eq:rpd}) is solved numerically for all AGB stars in the model clusters. In the dust-free case, the photodissociation radius is simply given by $r_p=\upsilon_e  / k^{\rm pd}_{\rm SiO}$.

The unattenuated photodissociation rate of SiO molecule through line absorption from the lower level $l$ into an upper state level $u$ is \citep{vanDishoeck:2006ec}
\begin{equation}
k^{\rm pd}_{\rm SiO} = \frac{\pi e^2}{m_e c} f_{ul} \eta_u \frac{4\pi J_\nu (\nu_{ul})}{h \nu_{ul} }, 
\end{equation}
where $f_{ul}$ is the line absorption oscillator strength, $\eta_u$ is the dissociation efficiency of the upper state $u$, which lies between 0 and 1. Assumption of $ \eta_u=1$ is reasonable for SiO molecule. Mutual shielding and self-shielding are the factors that can reduce the photodissociation rate of such molecules as CO, H$_2$. Mutual shielding is probably not important for SiO (E. van Dishoek, private communication). The total photodissociation rate is computed by summing over all lines. We include absorption to the 3~$^1\Sigma^+$ and 2, 3, 4, and 5~$^1\Pi$ states for SiO molecule with the oscillator strength $f=0.10$, 0.32, 0.03, 0.11, and 0.10, respectively, taken from the photodissociation database \footnote{\url{http://home.strw.leidenuniv.nl/~ewine/photo/}} \citep{vanDishoeck:2006ec}. The wavelengths for absorption lines are 1011, 1058, 1063, 1140, and 1378~\AA.

The mean intensity $J_\nu$ is derived from Eq.~(\ref{eq:MeanIntens}), for the intracluster radiation field, and is taken from work by \cite{Mathis:1983p5399}, for the ISRF.

\subsubsection{Wind model parameters}
We focus on dust formation by massive AGB stars with hot-bottom burning, therefore we consider only an oxygen-rich dust mixture in circumstellar envelopes. Although the dust composition in envelopes of M-stars depends on the initial stellar mass and mass-loss rates, for simplicity we assume a fixed silicate composition with density $ \rho_{\rm d}=3.3 \,\rm g\,\cmc$. We adopt the following values of the dust-to-gas ratio $D$ in the wind: $D=0$ corresponding to a dust-free case, and $D=0.001$, a typical value in outflows after the condensation zone derived from theoretical models of massive AGB stars. 
Grains are assumed to have a single size of 50~nm. The absorption efficiency $Q$ for a grain size of $a=50$~nm and $\lambda\sim100$~nm equates to 1. The optical depth at around 100~nm, calculated for the adopted grain size, is two times lower compared to that calculated for a power law distribution of grain sizes from work by \cite{Kim:1994iu}, which is commonly assumed in spectral energy distribution fitting.

The mass-loss rate is assumed to be isotropic. We adopt the mass-loss rate for oxygen-rich AGB stars derived empirically by \cite{VanLoon:2005ik}, which is more suitable for massive AGB stars than Reimer's law 
\begin{equation}
\label{eq:dmdt}
	\dot{M} = 1.38\times 10^{-11} \left( \frac{L}{L_{\sun}} \right)^{1.05}  
	\left( \frac{T_{\rm eff}}{3500\ \rm K} \right)^{-6.03},
\end{equation}
where $L$ and $T_{\rm eff}$ are the luminosity and effective temperatures of TP-AGB stars.
The expansion velocity $\upsilon_e$ is assumed to be 10\ \kms.
For Si abundance, the value $\epsilon_\mathrm{Si} = 4\times 10^{-5}$ is adopted. It is assumed that initially all Si atoms are bound in SiO molecules.

\begin{figure}[!tb]
    \centering
	 \includegraphics[width=0.5\textwidth]{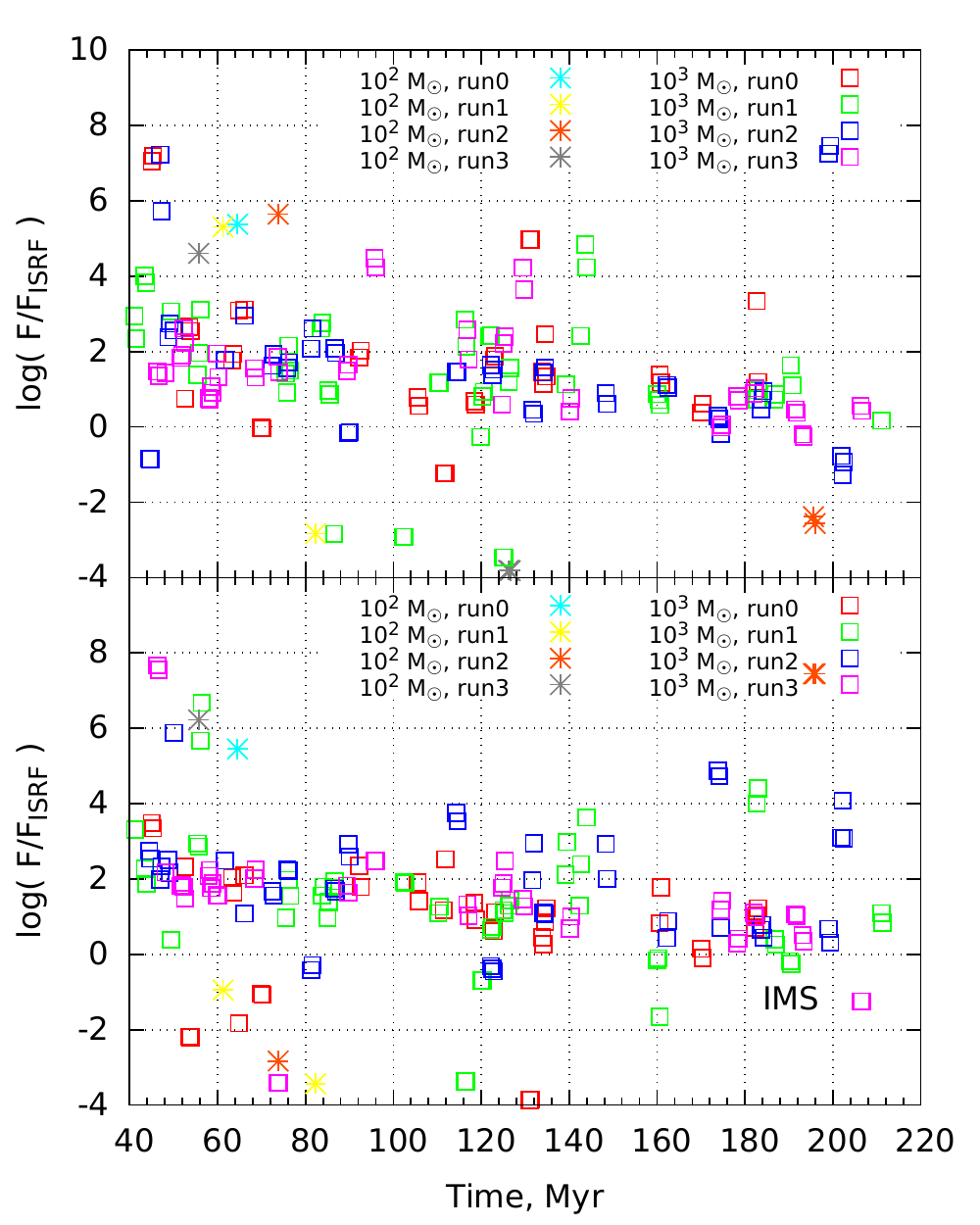}
	\caption{Intracluster UV radiation field in the locations of massive AGB stars in star clusters with initial mass of $10^2$ and $10^3\Ms$ relatively to the standard interstellar radiation field calculated for each simulation snapshot. Different colors show four simulation runs. Top and bottom panels show models without and with initial mass segregation, respectively.}
	\label{fig:irrad-agb-1}
\end{figure}

\begin{figure}[!tb]
    \centering
	  \includegraphics[width=0.5\textwidth]{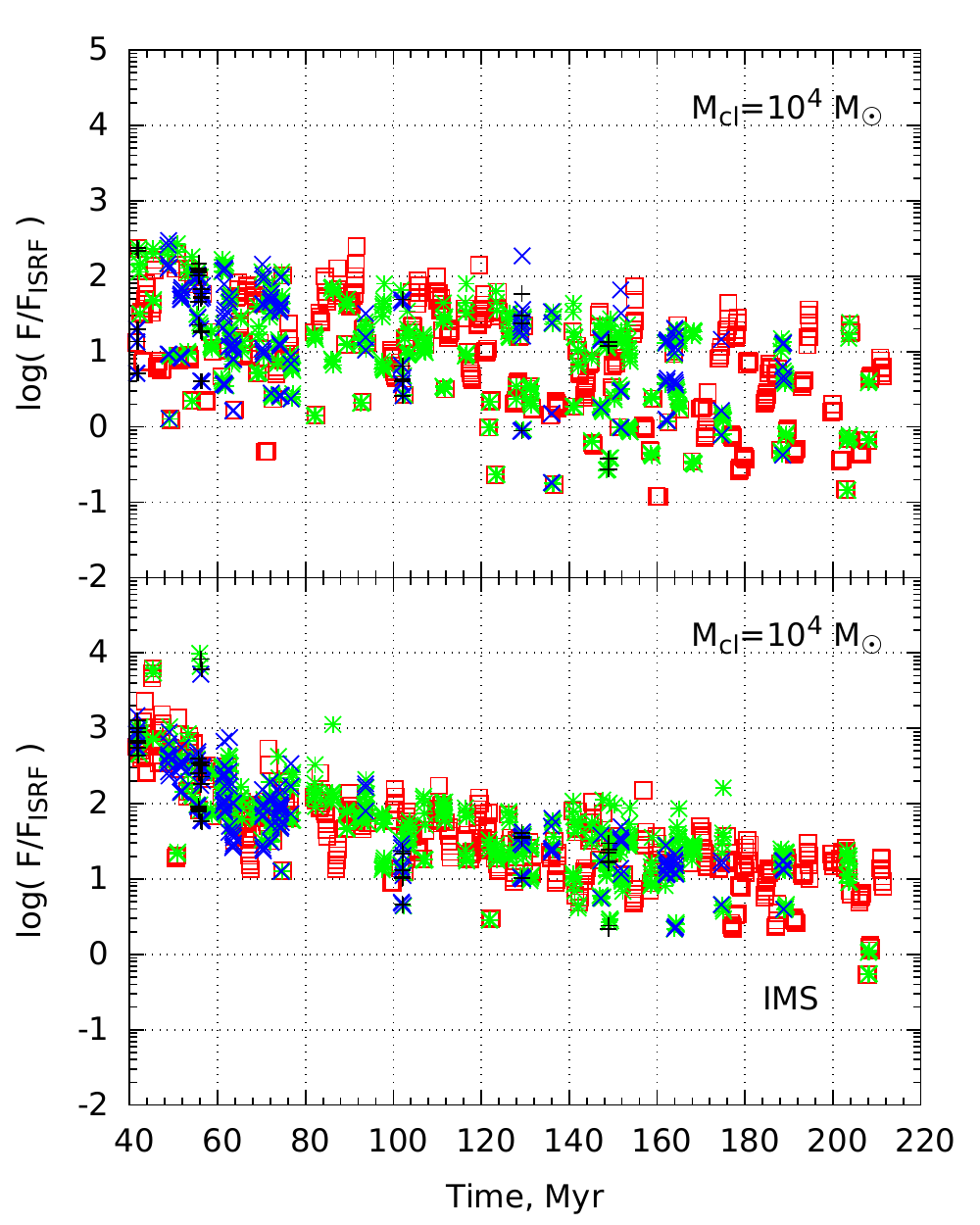}
	 \caption{Intracluster UV radiation field at the locations of massive AGB stars in star clusters with initial mass of $10^4\Ms$ without and with initial mass segregation (top and bottom, respectively) calculated for each simulation snapshot. The flux is shown relatively to the standard interstellar radiation field. Red squares, green stars, blue crosses and black plus symbols show UV field in the location of AGB stars in the snapshots with one, two, three and four AGB stars, respectively.}
	\label{fig:irrad-agb-2}
\end{figure}

\begin{figure}[!tb]
     \centering
	  \includegraphics[width=0.5\textwidth]{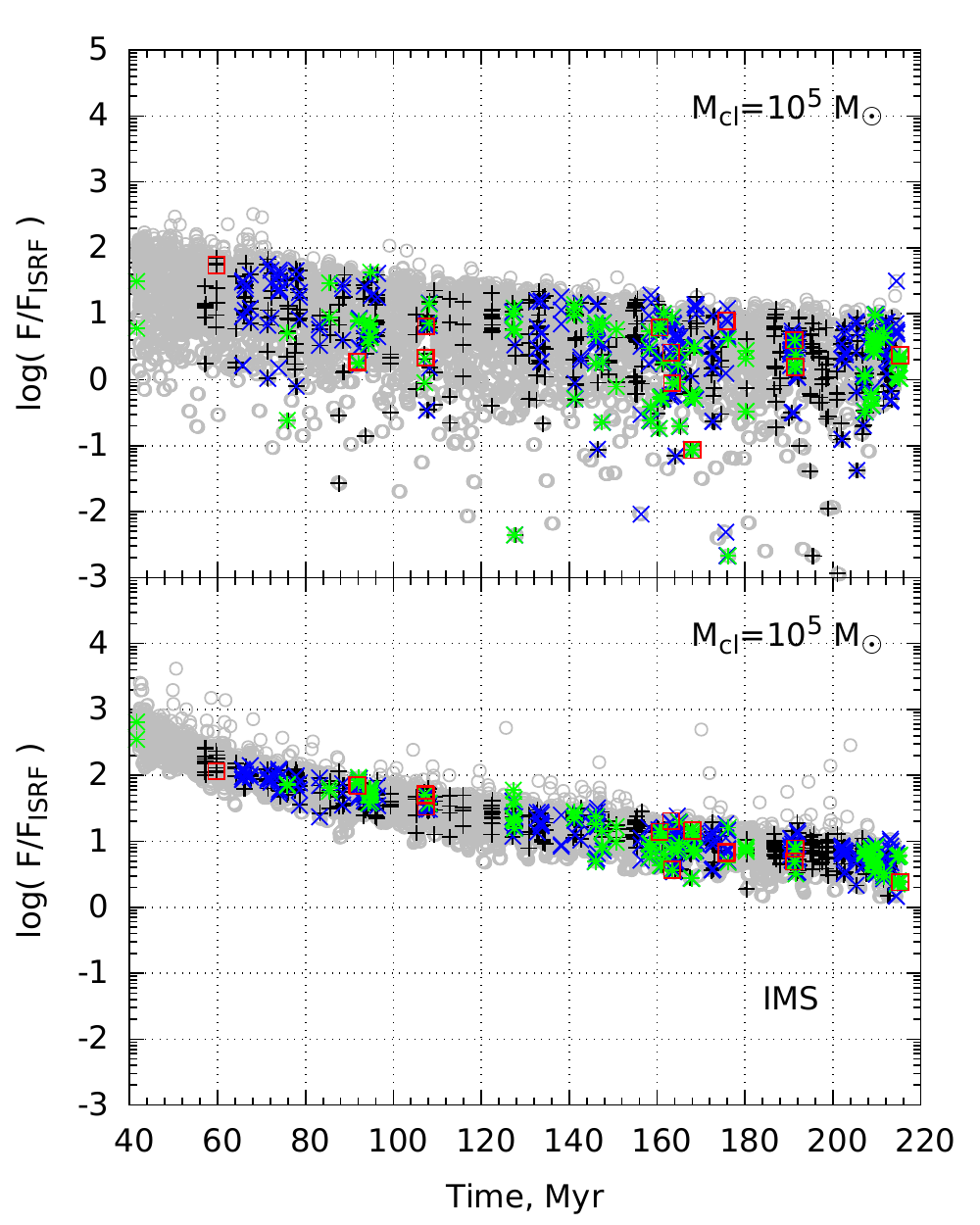}
	\caption{The same as in Fig.~\ref{fig:irrad-agb-2} for $M_{\rm cl}=10^5\Ms$. Grey circles indicate the cases, when there are more than four AGB stars in a snapshot.}
	\label{fig:irrad-agb-3}
\end{figure}




\begin{figure}[!tb]
\centering
	 \includegraphics[width=0.5\textwidth]{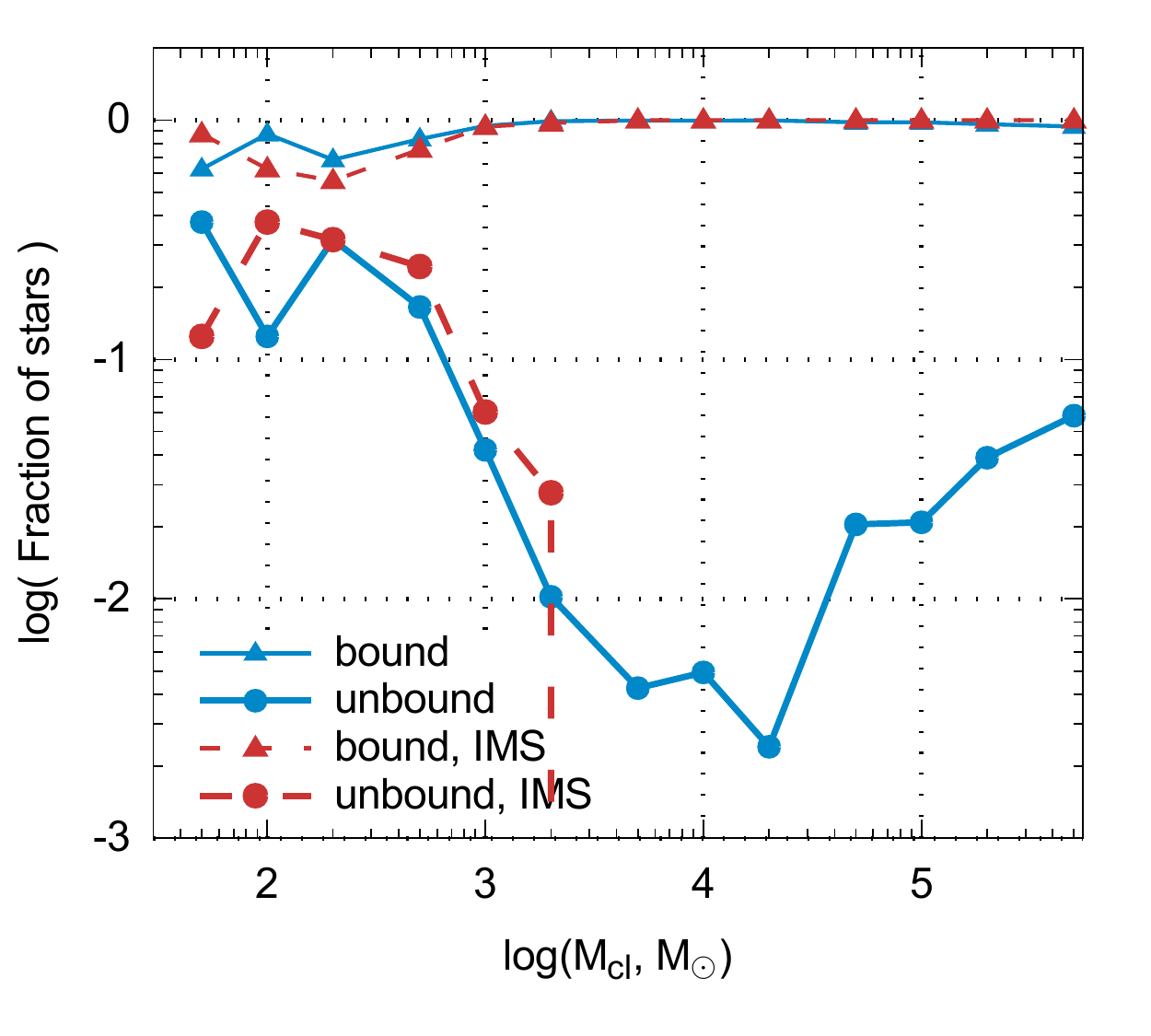}
\caption{Number of bound and unbound massive AGB stars (triangle and circle symbols, respectively) relatively to the number of these stars in each cluster as a function of initial cluster mass. Dashed and solid lines show star cluster models with and without IMS, respectively.}
\label{fig:AGBdistrCluster}
\end{figure}

\section{Results}\label{sec:results}
In this section, we present results of model evolution of massive AGB stars in star clusters of various initial mass  (Table~\ref{tab:InitialParamNbody}) rotating in the Galaxy at the solar galactocentric radius from simulation time $t=40$~Myr to $220$~Myr corresponding to the lifetimes of these stars. 

\begin{figure}[!tb]
    \centering
    \includegraphics[width=0.5\textwidth]{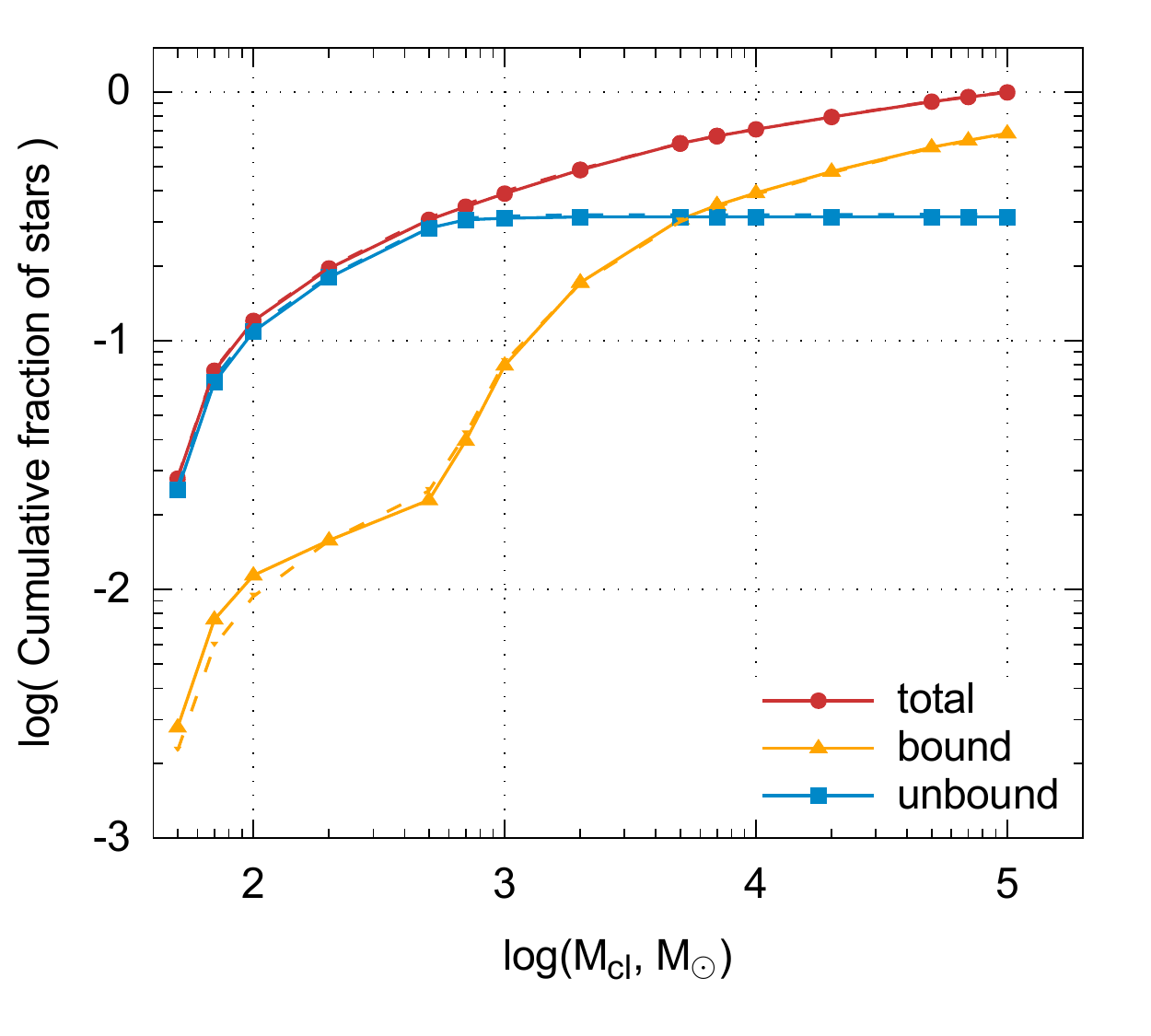}
\caption{Cumulative number of massive AGB stars evolving bound in the clusters, unbound (ejected) and their sum (triangles, circles and squares, respectively) relatively to the total number of these stars as a function of initial cluster mass $M_{\rm cl}$. Solid and dashed lines indicate results for models with and without initial mass segregation, respectively.}
\label{fig:AGBdistrCumul}
\end{figure}

\subsection{The fraction of massive AGB stars in clusters}\label{sec:res-AGBfrac}\label{sec:ResFracInClust}
The relative fractions of bound and unbound massive AGB stars for each model cluster are displayed in Fig.~\ref{fig:AGBdistrCluster}. The fraction is derived from the analysis of binding energies of all massive AGB stars in the cluster for the entire simulation run. Most of unbound massive AGB stars originate from low-mass star clusters. While low-mass clusters with initial mass segregation eject a slightly higher number of massive AGB stars compared to the models without IMS, it is the opposite for massive clusters ($M_{\rm cl}>2\times 10^3\Ms$) with IMS keeping almost all their massive AGB stars. The fractions of massive AGB stars lost by massive clusters modelled without IMS are 0.2--6\%.

Figure~\ref{fig:AGBdistrCumul} shows the cumulative distribution of massive AGB stars calculated with account of the mass distribution function of star clusters as described in Sect.~\ref{sec:ModFracAGBstars}. Most of unbound progenitors of massive AGB stars that eventually become members of the field population originate from star clusters with low-mass clusters, while more massive clusters preserve their massive AGB stars (see Fig.~\ref{fig:AGBdistrCluster}). The majority of low-mass clusters are dissolved in the process of emergence from their parent molecular clouds before the onset of our simulations. It is taken into account in the same way for both types of initial conditions, therefore the cumulative distributions derived  for models with and without initial mass segregation appear very similar in Fig.~\ref{fig:AGBdistrCumul}. Our main finding is that 70\% of massive AGB stars at the solar galactocentric radius reside in star clusters. 30\% of all massive AGB stars were lost from their clusters prior to the onset of AGB evolution and evolve in the field. They originate mainly from star clusters with initial mass $M_{\rm cl}<10^3\Ms$. In the following we focus on massive AGB stars in cluster environment and its effects on their circumstellar envelopes.

\subsection{Intracluster UV radiation field}
In the following, we investigate the strength of the intracluster radiation field, its time evolution and dependence on cluster mass using numerical simulations described in the Sect.~\ref{sec:SC-evolution}. 
Figures~\ref{fig:irrad-agb-1} -- \ref{fig:irrad-agb-3} show the strength of unattenuated intracluster radiation field in the locations of massive AGB stars for star clusters with initial mass $M_\mathrm{cl}=10^2,\ 10^3$, $10^4$, and $10^5\Ms$. The intracluster radiation field is calculated for each snapshot by summing up the fluxes of photons in the energy range of 5.6--13.598~eV from all cluster main sequence stars with mass $m \ge 2\Ms$. In order to compare the irradiation which an AGB star experiences in a star cluster to that in the ambient ISM, the fluxes are shown relatively to the standard interstellar radiation field (ISRF). We adopt the  value of the photon flux $F_{\rm Draine} = 1.921\times 10^8\, \cms\,s^{-1}$ \citep{Draine:1978p5783}. 

Since the duration of thermally pulsing AGB stage is relatively short, low-mass clusters have no more than one star in this evolutionary phase at the same time (Fig.~\ref{fig:irrad-agb-1}). A cluster with $M_\mathrm{cl}=10^4\Ms$ has up to four TP-AGB stars at the same time, while a $10^5\Ms$ cluster has generally more than four massive AGB stars. The UV fluxes at the location of multiple AGB stars are shown with different symbols depending on their number in Figs.~\ref{fig:irrad-agb-2} and \ref{fig:irrad-agb-3}. 

Although there is a scatter in the values of the UV flux, figures~\ref{fig:irrad-agb-1} and \ref{fig:irrad-agb-2} demonstrate that most of AGB stars are irradiated by 10-100 times stronger UV field than the ISRF. The average irradiation flux decreases with time as stars of higher mass go off  the main sequence (Fig.~\ref{fig:ion-flux}). There is more than an order of magnitude spread in UV flux strength towards lower values for clusters without initial mass segregation resulting from their more extended stellar distribution.

\begin{table}[tb]
\caption{Number of massive AGB stars with $R_{\rm pd, SiO} < R_{\rm cond}$ relatively to the number of these stars bound in SCs}
\label{tab:FracDustForm}
\vspace{3mm}
\centering
\begin{tabular}{c cccc}
\hline\hline
& \multicolumn{2}{c}{ IMS} & \multicolumn{2}{c}{no IMS} \\
\hline
$R_{\rm cond}/R_*$ & $D=0$ & $D=10^{-4}$ & $D=0$ & $D=10^{-4}$ \\
      2. &    0.60 &    0.04 &    0.40 &    0.05 \\
      5. &    0.80 &    0.25 &    0.60 &    0.14 \\
     10. &    0.92 &    0.64 &    0.75 &    0.41 \\
\hline
\end{tabular}
\\[2mm]
\textbf{Notes.} The first column gives the adopted values for the condensation radius $R_{\rm cond}$ relatively to the photospheric radius. It is followed by the relative numbers calculated for different values of the dust-to-gas ratio \textit{D} in the CSE for models with and without IMS.
\end{table}

\begin{table}[tb]
\caption{Number of massive AGB stars with $R_{\rm pd, SiO} < R_{\rm cond}$ relatively to the number of all massive AGB stars}
\label{tab:FracDustForm2}
\centering
\vspace{3mm}
\begin{tabular}{c cccc}
\hline\hline
& \multicolumn{2}{c}{ IMS} & \multicolumn{2}{c}{no IMS} \\
\hline
$R_{\rm cond}/R_*$ & $D=0$ & $D=10^{-4}$ & $D=0$ & $D=10^{-4}$ \\
      2. &    0.42 &    0.03 &    0.28 &    0.03 \\
      5. &    0.56 &    0.18 &    0.41 &    0.09 \\
     10. &    0.64 &    0.45 &    0.52 &    0.29 \\
\hline
\end{tabular}
\end{table}

\begin{figure*}[!t]
\centering
	\begin{tabular}{cc}
	\includegraphics[width=0.48\textwidth, page=1]{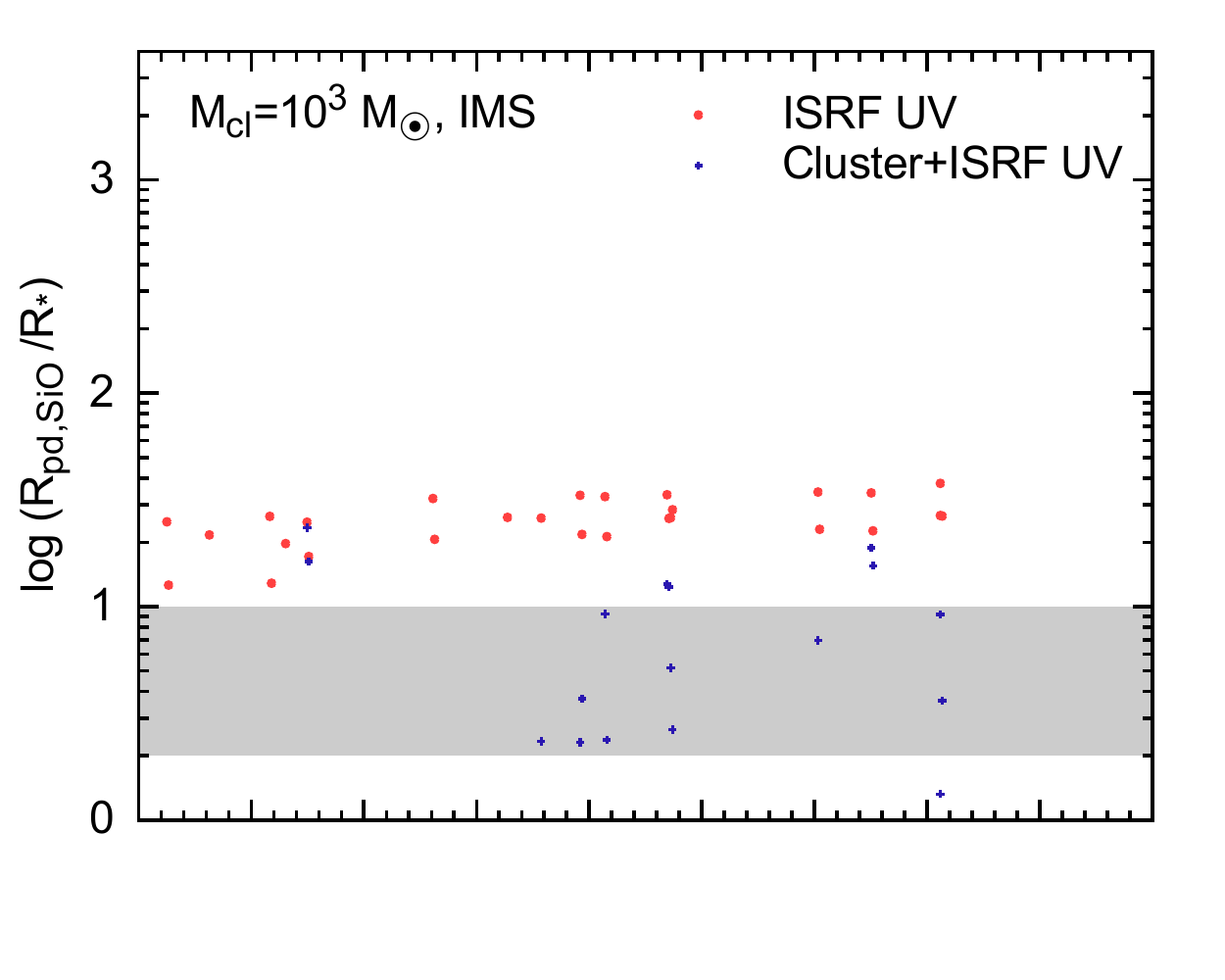}	&
	\includegraphics[width=0.48\textwidth, page=2]{fig8a} \\[-14.5mm]
	\includegraphics[width=0.48\textwidth, page=1]{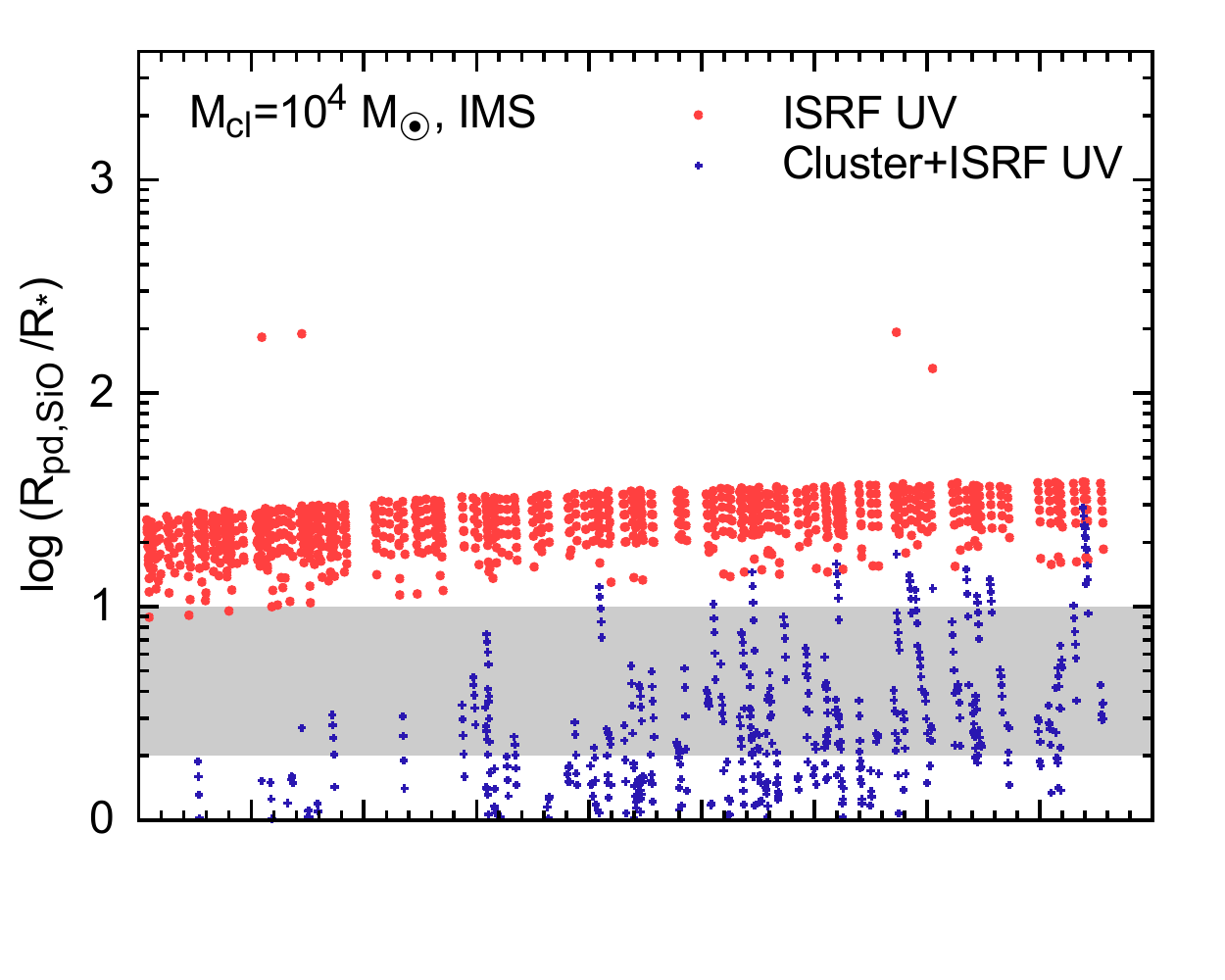}	&
	\includegraphics[width=0.48\textwidth, page=2]{fig8b}	\\[-14.5mm]
	\includegraphics[width=0.48\textwidth, page=1]{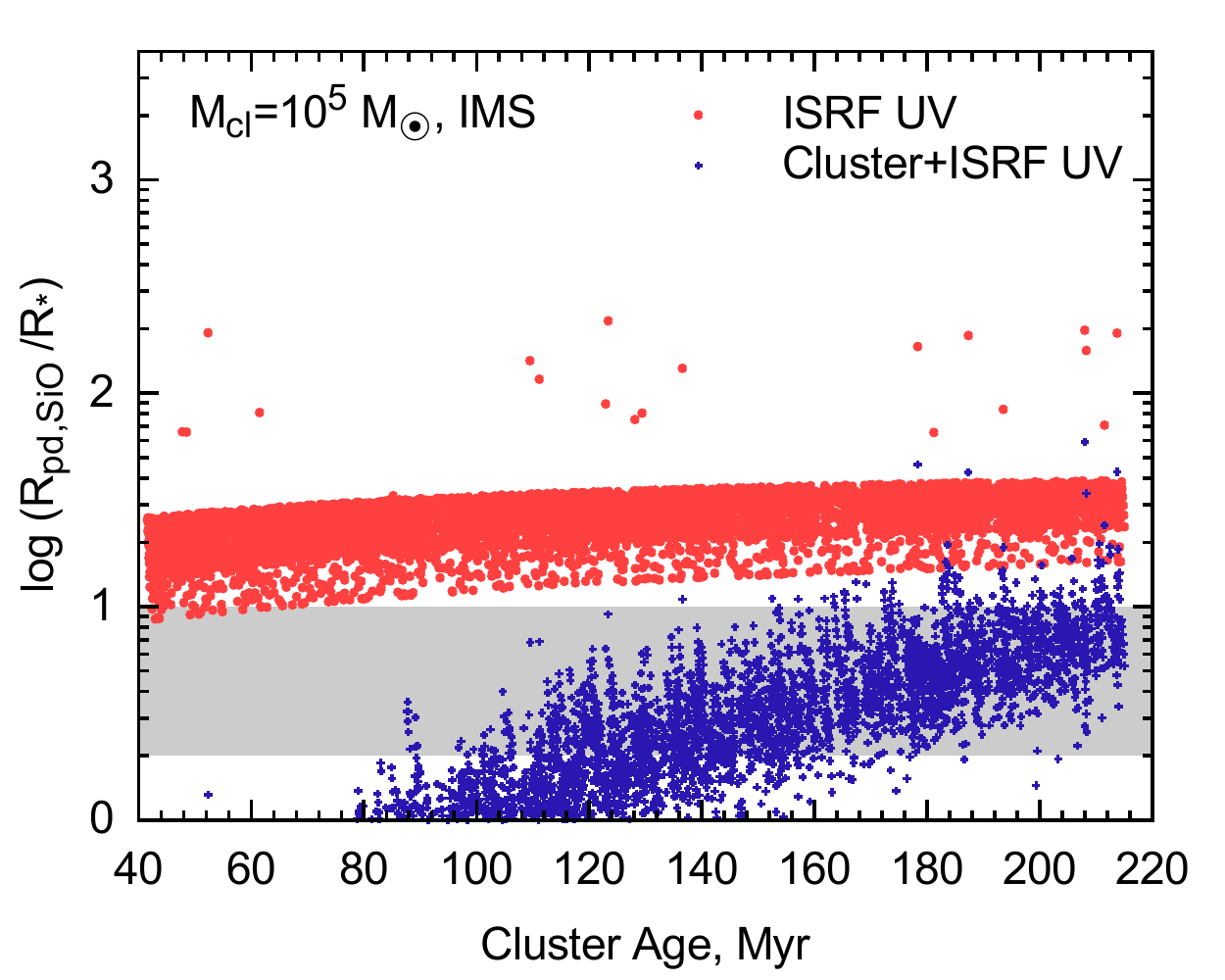}	&
	\includegraphics[width=0.48\textwidth, page=2]{fig8c}	
	\end{tabular}	
	\caption{Photodissociation radius of SiO relative to photospheric radius $R_*$ in dust-free shells of massive AGB stars in star clusters of $10^3$, $10^4$, and $10^5$ (from top to bottom). Red symbols show results of calculations with dissociation only by the ISRF component, and blue symbols by the ISRF and intracluster UV photons. Left and right panels present results for N-body models with and without IMS, respectively. The grey area marks the dust formation zone. }
	\label{fig:PDradius_d2g0}
\end{figure*}

\subsection{Photodissociation of SiO molecules by intracluster field}

\subsubsection{Photodissociation radii of SiO molecules in cluster AGB stars}
We compute the photodissociation radii $R_{\rm pd, SiO}$ of SiO envelopes of bound AGB stars in all simulated clusters for two cases of UV irradiation: (i) by the intracluster radiation field in the location of each AGB star in model clusters, and (ii) by the standard isotropic ISRF, for comparison. Figure~\ref{fig:PDradius_d2g0} shows the photodissociation radii for cluster masses of $10^3$, $10^4$, and $10^5 \Ms$ for MRR models with and without IMS for dust-free stellar wind. This case corresponds to the onset of dust condensation process. Time variations of the photodissociation radii of AGB stars in the cluster environment is similar to those of the intracluster UV field strength (Fig.~\ref{fig:irrad-agb-1}--\ref{fig:irrad-agb-3}). The difference is that the former is determined by the intensity in SiO dissociation lines, while the latter characterises the continuum UV radiation. 

During AGB phase, stars change their locations in the cluster, therefore they are irradiated by variable UV flux. From analysis of $R_{\rm pd, SiO}$ in different snapshots for the same AGB stars, we find that the corresponding variations of $R_{\rm pd, SiO}$ are typically less than 2 times.

It is insightful to compare the derived photodissociation radii with the dust condensation zone in CSE. The location of this zone is a matter of debate. According to theoretical models of dust condensation in outflows of M-stars, dust is formed at 5--10 photospheric stellar radii $R_*$ \citep[e.g.,][]{Gail:1999p2289, Ferrarotti:2003ho}. \cite{Nanni:2013wb} recently derived smaller radii for dust condensation of $2-3 R_*$ using similar to the aforementioned models with an alternative mechanism of dust destruction. A smaller condensation radius is also required by dynamical models of CSE of M-type AGB stars \citep{Hofner:2008et}. Observations of spatially resolved envelopes yield the values varying from a few to $\sim 10\ R_*$  \citep[e.g.,][]{Khouri:2014de, ZhaoGeisler:2012fi}. We highlight the radii from 2 to $10\ R_*$ in Fig.~\ref{fig:PDradius_d2g0} as the dust condensation zone.

As expected, $R_{\rm pd, SiO}$ due to photodissociation by the interstellar UV photons lies outside of the dust formation zone in all clusters even without dust shielding. The cluster UV photons penetrate much deeper in the CSE, reaching the dust condensation regions in a significant fraction of stars. A common feature for all model clusters is that the initial segregation of stellar mass has strong impact on $R_{\rm pd, SiO}$ owing to the higher stellar density of both irradiating and AGB stars in the central regions of clusters. 

\begin{figure*}[!t]
\centering
	\begin{tabular}{cc}
	\includegraphics[width=0.48\textwidth, page=1]{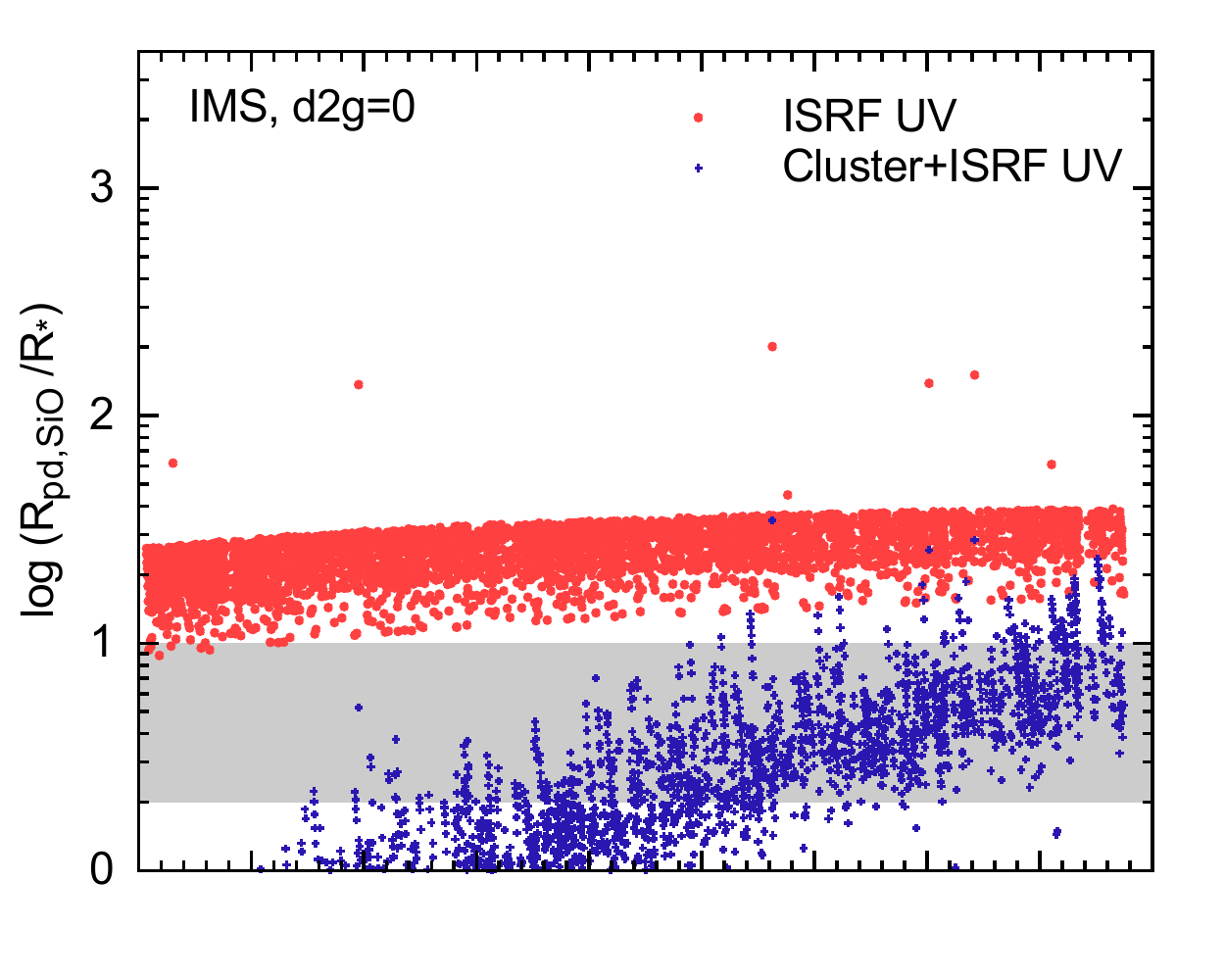}	&
	\includegraphics[width=0.48\textwidth, page=4]{fig9}	\\[-11.5mm]
	\includegraphics[width=0.48\textwidth, page=2]{fig9}	&
	\includegraphics[width=0.48\textwidth, page=5]{fig9}	\\[-11.5mm]
	\includegraphics[width=0.48\textwidth, page=3]{fig9}	&
	\includegraphics[width=0.48\textwidth, page=6]{fig9}	\\
	\end{tabular}	
	\caption{Photodissociation radius of SiO  relative to photospheric radius $R_*$  in massive AGB stars in a star cluster with mass of $5 \times10^4\Ms$ and dust-to-gas ratios of 0, $10^{-4}$, and $10^{-3}$ (top, middle and bottom panels, respectively). The meaning of the symbols is the same as in Fig.~\ref{fig:PDradius_d2g0}. Each second snapshot is shown to reduce crowding of the figure.}
	\label{fig:PDradius_vard2g}
\end{figure*}

Figure~\ref{fig:PDradius_vard2g} illustrates the dependence of the SiO photodissociation radii on the adopted dust-to-gas ratio $D$ in the shell. The figure shows  $R_{\rm pd, SiO}$ for stars in the $5\times 10^4\  \Ms$ cluster and the dust-to-gas ratio  $D=0$, $10^{-4}$, and $10^{-3}$. Similarly to other models, the intracluster radiation field in this cluster has the largest impact on SiO in the shells of more massive AGB stars ending their life during the first 100~Myr. In the dust-free case, the 54\% (18\%) of stars have  $R_{\rm pd, SiO} < 2R_*$ in models with(-out) IMS. For $D = 10^{-4}$, $R_{\rm pd, SiO}$ is noticeably pushed outside due to dust shielding. There are no stars with SiO envelope within  $2R_*$. Furthermore, 20\% of stars have $R_{\rm pd, SiO} < 5\ R_*$ in models with IMS and none without IMS. The fraction increases to 60\% (20\%) for our upper limit for the condensation radius of $10 \ R_*$ in the models with(-out) IMS.  The ratio $D=10^{-3}$ provides sufficient attenuation to block dissociating photons from the dust formation zone. The intracluster radiation in this case does not affect the condensation process, but it reduces the extent of SiO envelopes from a few times up to an order of magnitude. This effect can be used in observations of massive AGB stars as an indicator of cluster membership.

\subsubsection{Fractions of stars affected by the intracluster field}
We calculate the total number of bound AGB stars with $R_{\rm pd, SiO} < R_{\rm cond}$ for three values of the condensation radius $R_{\rm cond}=2,$ 5, and $10\ R_*$ discussed above for each simulated cluster. These numbers are used to derive the total fractions of massive AGB stars affected by the intracluster UV field by integration over the entire cluster mass range with the cluster mass distribution function. 
Table~\ref{tab:FracDustForm} shows the resulting fractions relatively to the total number of bound massive AGB stars. It is  derived using the total number of the bound stars $N^\mathrm{bound}_\mathrm{AGB}$ given by Eq.~(\ref{eq:NumBound}) in the denominator of Eq.~(\ref{eq:CumNumCorrected}) instead of $N^\mathrm{tot}_\mathrm{AGB}$. For $N^\mathrm{b}_\mathrm{AGB}$, we take the number of bound stars satisfying the condition $R_{\rm pd, SiO} < R_{\rm cond}$.
The fractions of stars  are  shown for models with and without IMS for the dust-to-gas ratio values $D=0$ and $10^{-4}$. 
For the dust-free case, $R_{\rm pd, SiO}$ in majority of AGB stars  in clusters with IMS is deeper than the inner border of dust condensation zone. The fractions of AGB stars in clusters affected by intracluster field are lower in models without IMS, but still significant, 60\% and 75\% with $R_{\rm pd, SiO} <5\ R_*$ and $R_{\rm pd, SiO} <10\ R_*$, respectively. The numbers in Table~\ref{tab:FracDustForm} imply that even if minor refractory dust species are formed closer to the star resulting in the dust-to-gas ratio of $\lesssim 10^{-4}$, shielding provided by these species is not sufficient and the intracluster UV field nevertheless may affect dust condensation in a large fraction of stars.

As discussed in Sect.~\ref{sec:res-AGBfrac}, results of our dynamical models of star cluster evolution combined with the adopted distribution of young star clusters imply that about 30\% of all massive AGB stars evolve in isolation. To illustrate the relative importance of cluster environment for the entire population of massive AGB stars (bound and unbound), we show the numbers of AGB stars with $R_{\rm pd, SiO} < R_{\rm cond}$  relatively to the number of all massive AGB stars in Table~\ref{tab:FracDustForm2}. According to our model calculations, in absence of dust shielding, the intracluster UV photons are able to penetrates in the CSE deeper than $10 R_*$ in 52 -- 64\% and deeper than  $2 R_*$ in 28 -- 42\% of all massive AGB stars. The lower and higher values correspond to the models without and with IMS.

\subsection{Implications for the stardust input from AGB stars}

\subsubsection{Revision of the model predictions for presolar grains}
In the following, we revise the model predictions for the relative contribution of massive AGB stars to the presolar grain inventory carried out in our earlier work \citep{Gail:2009p512}. \cite{Gail:2009p512} put forward a model, which relates theoretical studies of dust condensation during the entire AGB stage and a model for chemical evolution of dust and gas in the solar neighbourhood to the contribution of AGB stellar populations as a function of their initial mass and metallicity. This model suggests that 40\% and 60\% of silicate grains in the stardust population at the instant of solar system formation originate from low- and high-mass AGB stars, respectively.

In the present study, we can evaluate the maximum possible effect of cluster environment on the contribution from AGB stars by assuming that dust condensation is completely suppressed in massive AGB stars with $R_{\rm pd, SiO} < R_{\rm cond}$ and adopting the largest dust condensation radius $R_{\rm cond} = 10 R_*$. We use the same dust yields as in our earlier works \citep[e.g.,][]{Zhukovska:2008bw, Gail:2009p512, Zhukovska:2014ey} based on dust condensation models in O- and C-rich stellar winds of AGB stars. 
These yields are derived for single stars. There are observations that some binaries develop a common envelop which allows dust condensation process. Evolution of interacting binaries is poorly understood. There are presently no yields available for dust  condensed in the common envelopes  for the solar metallicity. Very recently, \cite{Zhu:2015fi} estimated the contribution of dust from binaries for the Large Magellanic Clouds and found that it is 4 times lower compared to the dust input from single stars using the dust yields from \cite{Zhukovska:2008bw}.

We multiply the dust masses ejected by massive AGB stars by the factor $1 - \eta$, where $\eta$ is the fraction of stars with photodissociation radii $R_{\rm pd, SiO} < R_{\rm cond}$ listed in Table~\ref{tab:FracDustForm2}. For $\eta=0.64$ taken for the IMS models, the corresponding revised value for the relative contribution of massive AGB stars to the O-rich presolar grains of the AGB origin equates to 0.4. We conclude that although including star cluster environment may reduce the relative contribution of massive AGB stars to the presolar grains, the lower limit for the mass fraction of 40\% is too high to help mitigate non-detection of grains from these stars in meteorites.

\subsubsection{Dust production rates}
AGB stars are the main stellar source of dust production in the Galaxy. The present day rate of dust injection from AGB stars at the solar galactocentric radius is $1.2  \times 10^{-3}\Mloss$, out of which $2.2 \times 10^{-4}\Mloss$ is attributed to the silicate stardust injection rate, accordingly to the models published in \cite{Zhukovska:2008bw}. The contribution from massive AGB stars is $8\times 10^{-5}\Mloss$ or 35\% of silicate dust input from AGB stars. Assuming that dust formation is completely suppressed in the cluster stars with $R_{\rm pd, SiO} < 10 R_*$ as described above and multiplying the dust yields for $m \geq 4\Ms$ by the reduction factor $1-\eta$ of 0.46, we attain a lower limit for the current injection rate for AGB stars of  $1.8 \times 10^{-4}\Mloss$ corrected for the fact that a substantial number of massive AGB stars remain bound in their parent star clusters. Thus including the effect of cluster environment in dust condensation models can reduce the total injection rates of silicate dust from AGB stars by at most 20\%.

In any case, if the total silicate production rates from AGB stars are decreased owing to the effects mentioned above, it should not affect the amount or composition of the interstellar dust mixture, since the mass fraction of stardust from AGB stars in the Milky Way is only 2\% and it is dominated by the carbonaceous grains. The majority of dust mass is formed by accretion in the ISM  \citep[e.g.,][]{Dwek:1998p67, Zhukovska:2008bw, Draine:2009p6616}.

\begin{table*}[tb]
\caption{Number of massive AGB stars with $R_{\rm pd, SiO} < R_{\rm cond}$ for the TRR models}
\label{tab:FracDustFormTRR}
\centering
\vspace{3mm}
\begin{tabular}{c cccc | cccc}
\hline\hline
& \multicolumn{4}{c}{ relatively to the number of these stars bound in SCs} & \multicolumn{4}{| c}{relatively to the number of all massive AGB stars}\\
& \multicolumn{2}{c}{IMS} & \multicolumn{2}{c}{no IMS}  & \multicolumn{2}{| c}{ IMS} & \multicolumn{2}{c}{no IMS} \\
\hline
$R_{\rm cond}/R_*$ & $D=0$ & $D=10^{-4}$ & $D=0$ & $D=10^{-4}$ & $D=0$ & $D=10^{-4}$ & $D=0$ & $D=10^{-4}$ \\
      2.  &    0.53 &    0.02 &    0.31 &    0.03 &      0.37 &    0.02 &    0.22 &    0.02 \\
      5.  &    0.74 &    0.21 &    0.51 &    0.09 &      0.53 &    0.15 &    0.36 &    0.06 \\
     10. &    0.89 &    0.57 &    0.70 &    0.34 &      0.63 &    0.40 &    0.49 &    0.24 \\
\hline
\end{tabular}
\end{table*}

\begin{table*}[tb]
\caption{Number of massive AGB stars with $R_{\rm pd, SiO} < R_{\rm cond}$ for the CRR models}
\label{tab:FracDustFormCRR}
\centering
\vspace{3mm}
\begin{tabular}{c cccc | cccc}
\hline\hline
& \multicolumn{4}{c}{ relatively to the number of these stars bound in SCs} & \multicolumn{4}{| c}{relatively to the number of all massive AGB stars}\\
& \multicolumn{2}{c}{ IMS} & \multicolumn{2}{c}{no IMS}  & \multicolumn{2}{| c}{ IMS} & \multicolumn{2}{c}{no IMS} \\
\hline
$R_{\rm cond}/R_*$ & $D=0$ & $D=10^{-4}$ & $D=0$ & $D=10^{-4}$ & $D=0$ & $D=10^{-4}$ & $D=0$ & $D=10^{-4}$ \\
      2. &    0.77 &    0.05 &    0.66 &    0.02 &    0.54 &    0.03 &    0.47 &    0.02\\
      5. &    0.86 &    0.60 &    0.78 &    0.42 &    0.60 &    0.42 &    0.56 &    0.30\\
     10. &    0.92 &    0.79 &    0.88 &    0.69 &    0.64 &    0.55 &    0.63 &    0.49\\
\hline
\end{tabular}
\end{table*}

\section{Dependence on the initial cluster mass--radius relation}\label{sec:dependIC}
In this section, we discuss how alternative relations between the initial mass and radius of model clusters based on the TRR and CRR models affects our main conclusions. All initial parameters for these models are the same as for the MRR set of models (Table~\ref{tab:InitialParamNbody}) with exception of the initial radii.

In all sets of models, the initial mass segregation allows to keep the majority of massive AGB stars bound in clusters with  mass above $ 2\times 10^3\Ms$. Without IMS, the MRR models for clusters with $M_{\rm cl} > 2\times 10^4\Ms$ are more affected by the tidal field than other models because of their largest initial sizes, but the number of ejected stars remains fairly low in all cases. The fraction of massive AGB stars evolving in clusters is 68\% (67\%) with (without) IMS for the TRR models, and 72\% (75\%) for the CRR models, respectively. These numbers are not very different from the value of 70\% for the MRR models.

The TRR models have a shallower slope of the mass--radius relation compared to the MRR models, 1/3 and 1/2, respectively. Therefore, the TRR models are characterised by lower intensity of intracluster UV radiation field in SCs with $M_{\rm cl} < 2 \times 10^4 \ \Ms $, and more intense radiation field in clusters of higher mass. The former decrease in the field strength is more important, because about 80~\% of progenitors of massive AGB stars originate from clusters with $M_{\rm cl} < 2 \times 10^4\  \Ms$. These results are quantified in Table~\ref{tab:FracDustFormTRR} showing the fraction of massive AGB stars in which the UV field penetrates in the dust formation zone normalised to the number of these stars bound in clusters and to their total (bound+unbound) number. The numbers of AGB stars with the radii of SiO photodissociation within the adopted values of $R_\mathrm{cond}$ for the TRR models are lower compared to the corresponding numbers for the MRR models (Tables~\ref{tab:FracDustForm} and \ref{tab:FracDustForm2}), but the difference is at most 10~\%.

The initial difference in the strength of the intracluster UV field compared the reference models is much stronger for the CRR case, although it is smoothed  to some extent by the tidal disruption of low-mass clusters during the simulations. In contrast to low-mass clusters, the sizes of massive clusters increase by less than 2 times at most and preserve the initial difference, which is about 5 times between the CRR and MRR models for the $10^5\Ms$ star cluster. Impact of these compact massive SCs on the total fraction of stars affected by the UV field is diminished by the cluster mass distribution function, proportional to $M_\mathrm{cl}^{-2}$ (Sect.~\ref{sec:ModFracAGBstars}). Nevertheless, the photodissociation radii of SiO are generally smaller in the CRR models, resulting in higher numbers of stars with $R_\mathrm{pd,SiO}<R_\mathrm{cond}$  (Table~\ref{tab:FracDustFormCRR}). In particular, the fraction of bound massive AGB stars with $R_\mathrm{pd,SiO}<5 R_*$ and $D=10^{-4}$ is 60\% (42\%) with(-out) IMS for the CRR models, which is more than two times higher than for the MRR models. The fraction of bound stars with SiO envelopes within 2~$R_*$, our most conservative value of the condensation radius, is by 17\% (25\%) higher in the CRR models with(-out) IMS and $D=0$ compared to the MRR.

This small dependence on the adopted initial mass-radius relation is attributed to the important role of the galactic tidal field in evolution of low- and intermediate-mass clusters. It implies that our conclusions do not strongly depend on the initial conditions.

\section{Discussion and concluding remarks}\label{sec:discussion}
We investigate what fraction of massive AGB stars evolves in their parent star clusters and impact of cluster UV radiation on their circumstellar environment. The field stars in our model originate (i) from low-mass clusters dissolve on the stage of emergence from the natal molecular clouds and (ii) by ejection from their natal star clusters. The latter process is modelled using N-body simulations of dynamical evolution of star clusters in the local Galaxy. We find that 70\% of massive AGB stars are members of their parent star clusters. This value can be lower if the fraction of stellar mass formed in star clusters is lower as suggested by some works \citep{Piskunov:2008ft, Kruijssen:2012bs}. Half of all bound massive AGB stars resides in clusters with initial mass exceeding $5 \times 10^3\ \Ms$. 

Intracluster UV radiation is $10^2-10^3$ times stronger than the interstellar radiation field for stars 6 -- 7~\Ms\ and gradually decreases to a few times ISRF strength for 4~\Ms\ stars in all considered star clusters. The UV field strength shows 10 times larger scatter towards lower values in clusters without initial mass segregation owing to more spatially extended stellar distribution. For this reason, the cluster UV photons on average penetrate deeper in the circumstellar shells of AGB stars in model clusters with initial mass segregation. 

We find that on the initial stages of dust formation the cluster UV photons are able to dissociate the SiO molecules in the dust formation zone in a large fraction of massive AGB stars. This implies that dust condensation process may be suppressed or delayed by intracluster UV field, similarly to the effect of UV field of chromospheric origin \citep{Beck:1992p6642}. The fraction of affected stars depends on the adopted initial mass--radius relation for model star clusters and location of dust-forming zone. The models with initial segregation of stellar mass are characterised by stronger UV fields and therefore smaller SiO envelopes.
 For our reference models, 80\% (60\%) of AGB stars bound in star clusters with (without) IMS have the SiO photodissociation radii within 5 photospheric radii. Relatively to all (bound and field) massive AGB stars, this fraction constitutes 56\% (41\%). This value is even higher, 60\% (56\%), if we assume the constant initial radius of clusters supported by some observations  \citep{Larsen:2004gf}.
Infrared surveys for dust-enshrouded giants in young star clusters in Magellanic Clouds point to lack of reddened IR objects  \citep[][]{VanLoon:2005fp}. However, they do find massive oxygen-rich AGB stars with dust emission signatures in some clusters. Our results imply that the sizes of SiO envelopes of AGB stars in cluster with ages $\lesssim 200$~Myr are smaller than those of their counterparts in the field population as seen in Fig.~\ref{fig:PDradius_vard2g}. This may be used as an indicator that an AGB star is a cluster member.  

Our findings have major implications for dust input from AGB stars at the solar galactocentric radius. Adopting the upper limit for the extent of condensation zone $R_\mathrm{cond}$ of 10~$R_*$ and assuming that dust formation is completely suppressed in stars with the SiO photodissociation radius $R_{\rm pd,SiO} < R_\mathrm{cond}$, we find that the effect of cluster environment on dust condensation in CSE can reduce the total injection rates of silicate dust from AGB stars from $2.2 \times 10^{-4}\Mloss$ to $1.8 \times 10^{-4}\Mloss$, or at most by 20\%. This value depends on the adopted dust yields and on the number ratio between AGB stars of low and intermediate mass, which is determined by the star formation history. Applying the same assumptions to the model of stardust lifecycle in the solar neighborhood, we derived the lower limit for the revised contribution of massive AGB stars to the presolar grain population of 40\%. This value is 1.5 times lower than our earlier estimate \citep{Gail:2009p512}, but it is still too high to help explain non-detection of grains with isotopic signatures of hot bottom burning in meteorites.

The lower mass limit for hot bottom burning adopted in the present work is 4~\Ms. If stars in mass range $\sim 3-4\ \Ms$\ also experience hot bottom burning \citep{Busso:1999p5975, Marigo:2007dc, Ventura:2012cs}, they will be less affected by intracluster UV field, since after 220~Myr its strength becomes comparable to or weaker than the ISRF. Consequently, the fractions of massive AGB stars affected by the intracluster field  would decrease, if we extend the considered mass range towards lower masses. 

In our simple estimate, we did not consider dissociation of molecules which may be precursors of minor dust species with higher condensation temperature, e.g. corundum ($\rm Al_2O_3$) and spinel (MgAl$_2$O$_4$) grains. We demonstrated that a low dust-to-gas ratio of $10^{-4}$ may mitigate the penetration of the UV photons in the base of dust formation zone. Minor oxygen-bearing dust species with higher condensation temperature can thus assist formation of silicate grains by blocking the UV irradiation. Clumpiness in the shell is another factor presently neglected in our study which can provide local shielding from UV photons and enable dust formation. 
 
Most mass is lost by evolved stars during a few last thermal pulses of AGB evolution (superwind phase). A possible implication of the results derived in this work is that the superwind stage may be delayed in most massive AGB stars in clusters owing to suppression of initial condensation by UV irradiation of CSE.

\section*{Acknowledgements}
SZ acknowledges support by the \textit{Deutsche Forschungsgemeinschaft} through SPP 1573: ``Physics of the Interstellar Medium''.
We are grateful to Hans-Peter Gail, Dmitry Semenov, Gael Rouill\'e, Stefan Schmeja and Diederik Kruijssen for fruitful discussions on various aspects of this study. We thank Peter Hoppe for reading the manuscript. We acknowledge the computation resources provided by the Rechenzentrum Garching and the GPU Cluster Milky Way at the Forschungszentrum J\"ulich.

\bibliography{AllReferences}

\end{document}